\documentclass[aps,pre,twocolumn,amsmath,amssymb,superscriptaddress,reprint,longbibliography]{revtex4-2}

\usepackage{siunitx}
\usepackage{booktabs}

\usepackage[utf8]{inputenc}
\usepackage[T1]{fontenc}
\usepackage[colorlinks=true,citecolor=blue]{hyperref}
\usepackage{graphicx}
\usepackage[nameinlink]{cleveref}

\begin{document}
\title{First-principles Calculations of MoSeTe/WSeTe Bilayers: Stability, Phonons, Electronic Band Offsets, and Rashba Splitting}
\author{Hamid Mehdipour}
\affiliation{Faculty of Physics, University of Duisburg-Essen, 47057 Duisburg, Germany}

\author{Peter Kratzer}
\affiliation{Faculty of Physics, University of Duisburg-Essen, 47057 Duisburg, Germany}
\email{peter.kratzer@uni-due.de}

\begin{abstract}
Janus materials have attracted much interest 
due to their intrinsic electric dipole moment and Rashba band splitting. 
We show that, by building bilayers of MoSeTe and WSeTe with different chalcogen atom sequences and different stacking patterns, one can modulate the net dipole moment strength and the Rashba effect, as well as the band alignment of the MoSeTe/WSeTe bilayer. Type-II band alignment is found which can be exploited to create long-lived interlayer excitons. 
Moreover, it is shown that the atomic sequence and stacking play pivotal roles in the interlayer distance of MoSeTe/WSeTe 
and thus its electronic structure and vibrational, especially low-frequency, characteristics. 
The long-range dispersion forces between atoms are treated with a conventional additive pairwise, as well as a many-body-dispersion method. 
It is shown that under the many-body dispersion method, more clear and rational thermodynamic trends of bilayer stacking are realized and interface distances are estimated more accurately. 
Vibrational spectra of the bilayers are calculated using first-principles phonon calculations and the fingerprints of monolayer attraction and repulsion are identified. An anti-correlation between distance and the shearing mode frequency of the rigid monolayers is demonstrated which agrees well with experimental findings.   
The results suggest that the judicious selection of the atomic sequence and stacking helps to widen the scope of the low-dimensional materials 
by adding or enhancing properties for specific applications, e.g. for spintronics or valleytronics devices.

\end{abstract}

\maketitle
\section{Introduction}
\par
Two-dimensional inorganic transition-metal dichalcogenide (TMDC) materials have attracted enormous interest over the past decade due to intriguing electronic and optical properties making them promising candidates for optoelectronic, sensing, and photocatalytic applications~\cite{mak2016photonics,samadi2018group,advmat_yun_20}. 
A tunable electronic bandgap, small dielectric constant, and intrinsic in-plane structural asymmetry play the determining roles in the functioning of such materials in the aforementioned applications. 
This in-plane symmetry breaking facilitates the exploitation of valley degrees of freedom for optical generation of valley-polarized electrons~\cite{ye_natnano2016electrical}. 
It was claimed that highly accurate spin manipulation becomes feasible in this way, paving the way to even more intriguing applications for this class of materials~\cite{xiaonatphys_14_valley}.

In addition to conventional di-selenides or di-tellurides, Janus materials, having different chalcogen species on either side of the transition metal atom, have widened the scope of TMDCs considerably. 
While for a long time the 
synthesis of Janus materials used to be merely an imagination, it turned into a reality using plasma etching~\cite{lu2017janus}, chemical vapor deposition \cite{schmeink2022} or pulsed-laser deposition~\cite{acsnano_gohegan_2020_janusplasma}. 
With Janus materials becoming available in high quality and larger quantities, one may also think of the next logical step and build stacks of {\em two different} Janus layers. 
This is the objective of the present theoretical study.

Already in conventional two-dimensional (2D) TMDC layers, 
breaking the out-of-plane inversion symmetry 
turned out to be an option for evoking new interesting electronic properties, thus broadening the range of TMDC applications~\cite{ye_natnano2016electrical,meng2018ferromagnetism}. 
For example, Rashba band splitting occurred after a sufficiently large charge density redistribution around the transition metal nuclei was evoked by a strong 
applied electric field normal to the monolayer~\cite{nanoscale_cheng_16-rashbafeild} or by assembling a bilayer of different TMDC monolayers~\cite{prb_lin_2019_rashbatmdc}. 

\par
Janus TMDC materials with different chalcogen atoms, MXY (M = Mo, W, and X, Y =  S, Se, Te), are thermodynamically less stable, but technically viable 
modified TMDCs that show additional 
intrinsic electrical and optical properties compared to the parent TMDC materials of $\text{MX}_2$~\cite{lu2017janus,wei2019second,zhang2017janus,le2017impact}. 
This is because the Janus TMDC monolayer intrinsically lacks the out-of-plane structural symmetry, as two different chalcogen atoms sit at mirror-symmetric lattice sites. The structural asymmetry and different electron affinities (of the chalcogen atoms) create a tangible electric dipole moment, resulting in a net electric field directed from an atom with lower  electronegativity to an atom with higher electronegativity~\cite{jpcc_tao_19_diplemoment_tuning,prb-mehdipour-22}. The electric field normal to the layer polarizes the electronic system~\cite{lu2017janus}, 
leading to an observable spin Hall conductivity in valence bands~\cite{prb_yu_21_rashbasoc}. The broken-symmetry-induced electric field also enhances the nonlinear optical response of layers~\cite{wei2019second}. The strength of electric charge polarization is very sensitive to the stoichiometry of MXY monolayers and their possible stacking (up to the bulk counterpart of monolayers). This area of research on the Janus materials has been explored 
using real-time  first-principles methods to compute the electronic structure and nonlinear susceptibility of Janus monolayers of TMDCs with different stoichiometries of the chalcogens~\cite{lucking2018large} and different stacking patterns of the same monolayers~\cite{wei2019second,zhang2017janus}. 

\par
In the present work, we suggest combining {\em two different} Janus materials into a bilayer heterostructure and carrying out first-principles calculations for such systems. 
The numerous structural variants, differing both in their chalcogen atom sequence and in the registry between the upper and lower  layer, offer many more possibilities to tune the band alignment and local electric field, as well as 
symmetry-breaking deformation of orbitals, 
which may influence the Rashba band splitting.
For applications in spin transistors, a large Rashba effect is desirable, which led us to select the heavier chalcogens, Se and Te, for our study. 
By modulating the asymmetry-induced net electric field of Janus monolayers one can adjust the newly emerged properties easily to suit specific applications. Yao {\it et al.} showed that a small in-plane biaxial strain can be used to fine-tune the Rashba effect ~\cite{prb-yao-17} in a WSSe monolayer. In fact, an applied in-plane strain can modify the W-Se bonding interaction, i.e., the orbital overlap and the charge accumulation at different sides of the layer.
Eventually, a change in the overlap of orbitals leads to a change in Rashba parameter for the valence band maximum. 

A few theoretical papers already addressed Janus bilayers. 
Wang {\it et al.} \cite{jphyconmat-Wang-2019} studied the effect of the interlayer potential drop and interlayer interactions on the band offset as well as the band gap nature for WSSe/MoSSe  heterostructures with different stacking sequence and interface pattern. This study showed that the Rashba parameters, important for spintronic applications, depend strongly on the interlayer and intralayer potential drop, i.e. the stacking and chalcogen atom sequence in the WSSe/MoSSe band structure.
Theoretical work by Li {\it et al.}~\cite{jpclLi17_mowsse} showed that in-plane Rashba polarization can be enhanced by constructing a type-II  heterostructure of WSSe/MoSSe due to improved out-of-plane electric polarity. 
Using the first-principles method, Gua {\it et al.} \cite{pccp-gua-20} investigated the structural, mechanical, and electronic properties of the MoSSe/WSSe van der Waals (vdW) heterostructure under various degrees of horizontal and vertical strain. 
However, less attention has been paid to the way in which  the atomic sequence and the registry between subsequent layers  affect  the most important electric properties, such as the Rashba effect, as well as the vibrational spectrum of such polar heterostructures.
So far, only for multilayers of a single TMDC material low-frequency shear vibration modes 
have been observed~\cite{apl-Plechinger-12}. Puretzky {\it et al.}~\cite{acsnano-puretzky-15} detected shear and layer-breathing rigid vibrations of monolayers in homo- and hetero-bilayer TMDCs. A strong dependence of vibration frequencies on the types of junction and stacking orientation was observed and supported by several follow-up studies~\cite{apl-elis-11,apl-holler-20,apl-ufuk-22} on similar bilayer structures that may even include Janus monolayers.

\par
In this paper, we perform a comprehensive first-principles investigation of MoSeTe/WSeTe bilayers, including four different chalcogen sequences with six possible stacking patterns each. Our calculated vibrational modes allow for detailed comparison with experimental data and thus put the experimentalists in position to characterize their samples e.g. via Raman spectroscopy. 
Moreover, the vibrational analysis demonstrates the effect of the inter-layer coupling strength modulated by the chalcogen atom sequence on the very-low-frequency spectrum of the MoSeTe/WSeTe bilayer. 
We calculate the electronic structure with a hybrid density functional that yields realistic values for the electronic band gap. 
Our results suggest that the sequence of chalcogen atoms in the MoSeTe/WSeTe bilayer plays a pivotal role in determining the electronic properties of the MoSeTe/WSeTe bilayer, such as the type of band alignment and the size of the electronic band gap, the monolayer contribution to the electronic band edges, the magnitude of the conduction band offset, as well as the band splitting at the high-symmetry point due to spin-orbit-coupling caused by the heavy atoms and the lack of inversion symmetry. 
In the case of Rashba-type splitting, we show that the stacking orientation can be used to tune the Rashba parameters. 

\section{Computational Methodology}
\paragraph{Density Functional Theory} 
We study the thermodynamic stability of possible bilayers of Janus MoXY and WXY with X = Se, and Te. The electronic structure and vibrational spectra of all the bilayer geometries are also calculated and discussed. All the relaxation and post-processing electronic structure calculations were carried out using density functional theory (DFT) as implemented in the FHI-aims code~\cite{BLUM20092175}. This code provides an all-electron description in which each single-particle Kohn-Sham (KS) electronic orbital is described by a linear superposition of several numeric atom-centered basis functions.

\par
To treat electronic exchange and correlation, the generalized-gradient approximation (GGA) parameterized as PBE~\cite{perdew_PhysRevB.46.6671} is used. 
However, for obtaining electronic band structures, the hybrid functional HSE06 is used on top of the PBE-relaxed geometry to obtain more realistic band gaps.
The Broyden-Fletcher-Goldfarb-Shanno~\cite{Fletcher87} method is employed to relax the atomic positions, and the maximum residual force component per atom was set at $5.0 \times 10^{-3} \text{eV/\AA}$. 
For k-point integration over the first Brillouin zone, an energy-converged Monkhorst-Pack k-mesh of $15\times15\times1$ has been used. 
Periodic boundary conditions are combined with a vacuum space of nearly $15~\text{\AA}$ perpendicular to the layer to avoid any artificial interactions between the bilayer and its image replica.

The so-called 'tight' basis set of the FHI-aims code is employed to carry out the relaxation of atomic positions. To account for relativistic effects on the electronic  band structures, spin-orbit coupling (SOC) is taken into account by performing a second-variational SOC calculation, as a post-processing step with $2N$ KS states obtained in the last self-consistent field iteration~\cite{blum_PhysRevMaterials.1.033803}.
Since each Janus monolayer is a polar material (due to the different electronegativity of the chalcogen atoms), and thus the bilayer is polar as well, the dipole correction method~\cite{prb_neugebauer_20} is used to eliminate the spurious electrostatic interaction between the bilayer and the image replica normal to the bilayer plane. By nullifying the electric field in the center of the vacuum region, one can estimate the induced potential jump, which is proportional to the total electric dipole moment density $D_{\text{tot}}$ of the bilayer. By calculating $D_{\text{tot}}$ for each bilayer, the effects of different chalcogen ordering and layer stacking are investigated and discussed.

\paragraph{Thermodynamic stability} 
To examine the relative (interfacial) stability of every heterostructure, the binding energy $E_b$ is calculated using 
\begin{equation}
E_b = E_{\text{MoXY/WXY}}-E_{\text{MoXY}}-E_{\text{WXY}},
\end{equation}
where $E_{\text{tot}}$ is the total energy of the heterostructure, and $E_{\text{MoXY}}$ and $E_{\text{WXY}}$ are total energies of the isolated Janus MoXY and WXY monolayers, respectively. 

\paragraph{Dispersion forces: Additive Pairwise vdW versus Many-Body Dispersion Interaction}
In the present work, we use two methods to account for the weak van der Waals interaction between the monolayers: a) an additive pair-wise  inter-atomic potential, and b) a many-body dispersion approach. 
Since we encountered significant structural changes between a) and b), the two methods are briefly presented and discussed. Technically speaking, the vdW interaction is a non-local contribution to the electronic correlation energy $E_c$. Both methods can be derived from the adiabatic-connection fluctuation-dissipation (ACFD) formalism~\cite{prl_Toulouse_09}, but with different levels of approximation.  The ADFC theorem states that 
 \begin{equation}
     E_c = - \frac{1}{2\pi}\int^{\infty}_{0} d\omega~\int^{1}_{0}~d\lambda\text{Tr}\left[\left(\chi_{\lambda}(\omega) - \chi_0(\omega)  \right) v \right] \, , 
     \label{eq-corr}
 \end{equation}
involving integration both over frequency $\omega$ and a running coupling constant $\lambda$. The quantities $\chi_{\lambda=1}(\omega)$ and $\chi_0(\omega)$  are the susceptibilities of the fully interacting and non-interacting Kohn-Sham electronic system, respectively. 
After performing $\lambda$-integration, the results can either be expressed in terms of $\chi_0$, 
 \begin{equation}
     E_c = \frac{1}{2\pi}\int^{\infty}_{0} d\omega   \text{Tr}\left[ \ln(1 - \chi_{0} v) + \chi_0 v  \right] \, ,
     \label{eq-corr-chi0} 
 \end{equation}
or be expanded in powers of $\chi_1$
 \begin{equation}
     E_c = - \frac{1}{2\pi}\int^{\infty}_{0} d\omega~\sum^{\infty}_{n = 2}(-1)^{n}\left(1 - \frac{1}{n}\right) \text{Tr}\left[\left(\chi_{1} v\right)^{n}\right] \, .
     \label{eq-corr-fpa} 
 \end{equation}
For deriving  an additive pair-wise  inter-atomic potential, only the leading term, being equal  in both expansions, matters. 
Specifically, we use the Tkatchenko-Scheffler (TS)\cite{prl_tkatchenko_09} form of the interaction potential,
\begin{equation}
    E_{\rm TS}^{\text{vdW}} = -\frac{1}{2} \sum_{A,B} f_{\text{damp}}\left(R_{AB}, R_{A}^{0}, R_{B}^{0}\right)C_{6AB}/R_{AB}^{6},
    \label{eq-evdw}
\end{equation}
 where $R_{A}^{0}$ and $R_{B}^{0}$ are vdW radii of atoms $A$ and $B$, and $R_{AB}$ is the distance between these atoms. $f_{\text{damp}}$ is a short-ranged damping function that eliminates the singularity originating from the interactions between very close atoms. In the $C_{6AB}$ coefficients, the  (usually rather mild) dependence on the ground-state electron density around each nucleus enters via a Hirshfeld partitioning scheme. Hence, the TS method allows one to consider the bonding environment of the atoms only in a superficial way.
 
 An improved description of environmental effects on the vdW interaction is obtained by treating each atom as a harmonic oscillator, with all atomic oscillators interacting via their fluctuating dipole fields. The so-defined MBD method \cite{prl_tk_car_12,jcp_tkatchenko_13} that we use here 
 allows us to work with the full interacting susceptibility $\chi_1$ in eq.~(\ref{eq-corr-fpa}) 
 as long as we restrict ourselves to interactions of the dipolar form.  
Technically speaking, $E_{\text{vdW}}$ is calculated as 
 \begin{equation}
     E_{\text{vdW}}  = %E_{\rm TS}^{\text{vdW}}  + 
     \frac{1}{2\pi} \int^{\infty}_{0} d\omega  \ln  \text{Det}\left[ C^{\rm RPA}(i \omega)  \right] \, ,
     \label{eq-corr-CRPA} 
 \end{equation}
where $C^{\rm RPA}$ is an interaction matrix resulting from the exact diagonalization of a coupled-oscillator model. This enables the inclusion of non-local interactions over distances of several atomic bond lengths and thus (in many cases) leads to an improved description of environmental effects.

The vdW interaction treated via the MBD scheme is generally found to be less attractive than the interaction described by the TS pair-wise potentials. 
This can be understood from analyzing the higher-order contributions to the expansion (\ref{eq-corr-fpa}): 
While a series expansion of the logarithm in (\ref{eq-corr-chi0}) yields all terms with the same sign, 
in eq.~(\ref{eq-corr-fpa}) the terms with even powers $n$ contribute with negative sign~\cite{jcp_tkatchenko_13}, leading to a less attractive overall interaction.

\section{Results and Discussion}  \label{Performance}
\subsection{Thermodynamic Stability of MoSeTe/WSeTe Bilayers: Effect of Stacking and Dipole Moment}

After first optimizing the lattice constants of each Janus layer separately and then taking the arithmetic average,  
the atomic positions 
were relaxed for both MoSeTe and WSeTe. Total energy calculations were performed for the relaxed geometries at a converged k-mesh size. The total energy difference at the 
materials-specific versus average lattice constants 
appears to fall below meV/formula unit, thus one may conclude that each monolayer undergoes a minimal strain when it is brought in contact with the other. This is because the lattice constant of a Janus monolayer is governed by the size of chalcogen atoms, Se and Te, rather than that of the transition metals. Also, we should underline that the lattice constants remain unchanged upon changing the chalcogen atomic sequence and bilayer stacking. 
By enumeration one finds 
four possible bilayer atomic junctions, Te1-Se2, Te1-Te2, Se1-Se2, and Se1-Te2. 
The interlayer distance, the junction gap, is a defining characteristic of a junction. Of course, the larger the junction atoms, the larger the distance. However, it will be further modified by the attractive or repulsive forces due to the inhomogeneous charge distribution within each layer and due to the fluctuating dipolar forces responsible for the vdW interaction. 

\begin{figure}[ht]
\centering
\includegraphics[width=0.53\textwidth]{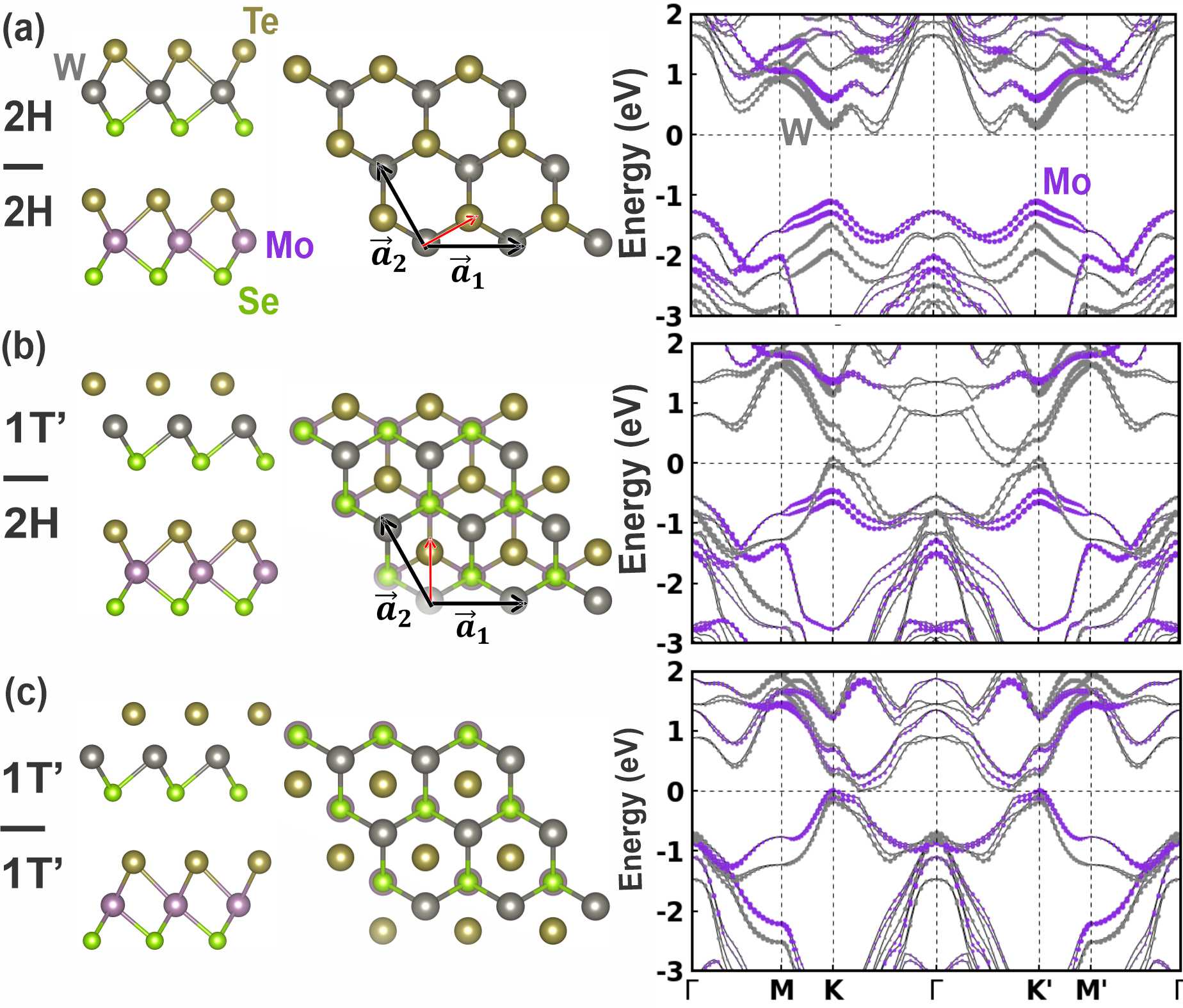}
\caption{Side (left column) and top (middle column) views, and electronic band structures (right column) for stacking of different phases of Janus MoSeTe and WSeTe monolayers: (a) 2H/2H, (b) 2H/1T$'$, and (c) 1T$'$/1T$'$ phase combinations. The red arrows in (a) and (b) show how the 1T$'$ phase may form by $60^{\circ}$ rotation around the $z$ direction.}
\label{fig-bilayer-different-phases}
\end{figure}

Experimentally, high-temperature growth techniques, e.g. chemical vapor deposition~\cite{acsnano-puretzky-15}, are typically employed to fabricate the monolayers of Janus materials that form the building blocks for the bilayers that we investigate computationally. 
Therefore, we start our computational investigations by exploring the thermodynamic stability of the Janus monolayers MoSeTe and WSeTe. 
The Se and the Te atoms of one Janus layer may be placed either on staggered or eclipsed positions (see Fig.~\ref{fig-bilayer-different-phases}), leading to the 2H or 1T' phase, respectively. 
From a theoretical perspective, the 1T$'$ phase of an isolated WSeTe monolayer can be constructed starting from 2H by rotating the Te atomic plane by $60^{\circ}$ around the $z$ direction, see Fig~\ref{fig-bilayer-different-phases}.
According to our PBE calculations, the MoSeTe Janus monolayer is found to be energetically most stable in the 2H phase, while WSeTe crystallizes more likely in 1T$'$ phase. 
This is proven by the fact that the 2H phases of MoSeTe and WSeTe are $76$ more, and $380~\text{meV}$ per unit cell less stable than the 1T' phase, respectively. 

Moreover, Fig.~\ref{fig-bilayer-different-phases} shows three structural models of MoSeTe/WSeTe bilayers constructed by combining the two phases the constituting Janus monolayers 
 can form. One may conceive that a 2H-MoSeTe/1T$'$-WSeTe bilayer renders the most stable heterostructure, but contrary to expectations,  our total energy calculations, employing the MBD method for the dispersion interactions, show that 2H-MoSeTe/2H-WSeTe bilayer is by $550~\text{meV}$ per unit cell more stable than the  2H-MoSeTe/1T$'$-WSeTe bilayer. This suggests that an exothermic transformation of the 2H-MoSeTe/1T$'$-WSeTe bilayer to a 2H-MoSeTe/2H-WSeTe bilayer (through atomic rearrangement within the topmost chalcogen layer of the WSeTe) is likely at room temperature. Also, 1T$'$-WSeTe has metallic character,  
 see Fig.~\ref{fig-bilayer-different-phases} (b) and (c). 
 Since we have applications for semiconductor heterostructures in mind, we discard the bilayers containing 1T$'$-WSeTe and only consider the most stable 2H/2H structure in the remainder of this paper.

% This figure shows bilayers with different phases and their corresponding band structure

\par
\begin{figure*}[ht]
\centering
\includegraphics[width=1.000\textwidth]{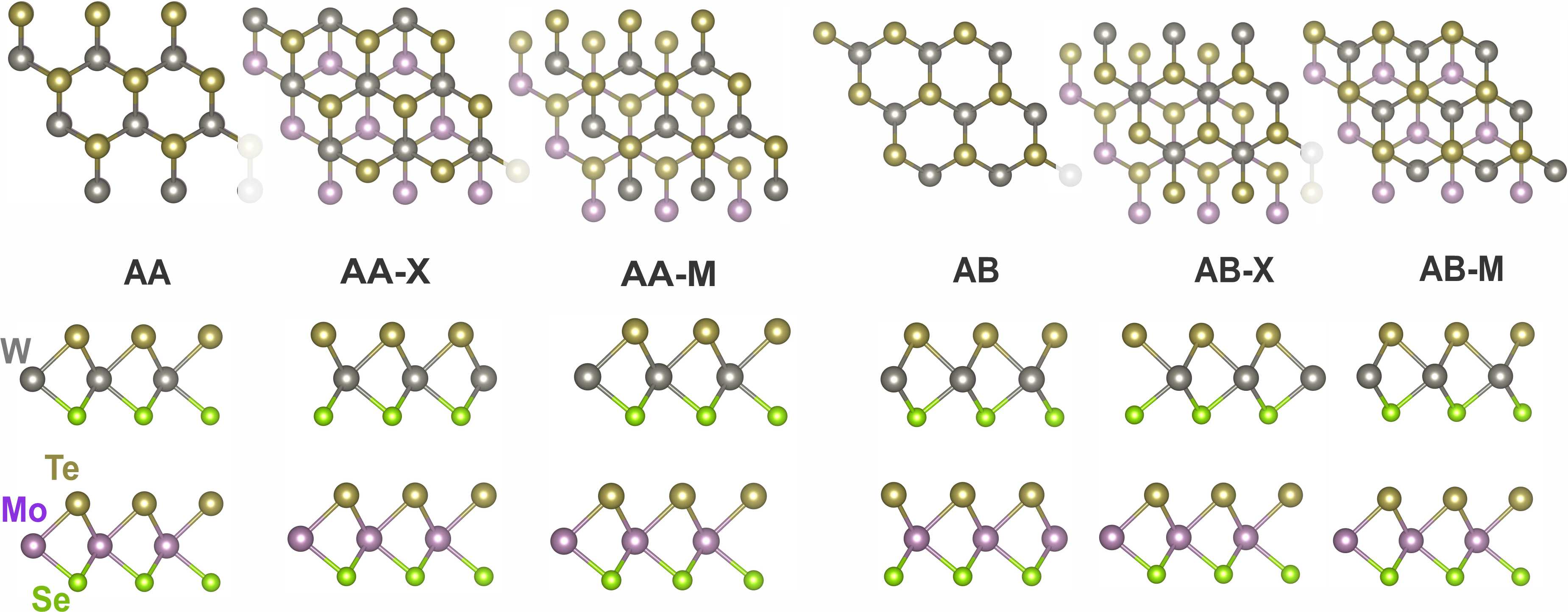}
\caption{Top (top row) and side (bottom row) views of different stacking patterns of MoSeTe/WSeTe bilayer. The two classes of stacking are parallel (AA) and anti-parallel (AB), which are subclassified based on chalcogen (-X) or transition metal (-M) atom of the top layer sitting in the hollow site of the bottom layer.}
\label{fig-sketch}
\end{figure*}

Figure~\ref{fig-sketch} displays hexagonal ($1\times1$) supercells of the MoXY/WXY bilayers constituted by the transition metals Mo and W and the two chalcogens X and Y (Se and Te). Supercells of heterostructures MoXY/WXY with four possible sequences of chalcogen atoms: SeMoTeSeWTe, SeMoTeTeWSe, TeMoSeSeWTe, and TeMoSeTeWSe are constructed and named CAS-1, CAS-2, CAS-3, and CAS-4, respectively. 
We note that for CAS-2 and CAS-3, the junction between the two monolayers is formed by the same chalcogen species, whereas CAS-1 and CAS-4 have junctions of Se and Te atoms pointing toward each other. 
For each atomic sequence, six stacking patterns are realized based on the two transition metal basal planes pointing in the same (AA) or different (AB) direction 
and either the chalcogen (-X) or transition metal (-M) atom of the top monolayer sitting at the hollow site of the bottom monolayer, see Fig.~\ref{fig-sketch}. In the literature, the AA and AB sets of stacking are referred to as the R- and H-type of stacking orientations. 

\par
Calculating the binding energy of the bilayer provides a measure of the strength of the vdW interaction between the monolayers (MLs). This aids us to estimate the impact of chalcogen sequence and stacking pattern on the vdW contribution to the bilayer binding. Moreover, for a hetero-bilayer composed of different Janus MLs with non-zero electric dipole moments, the binding energy can be used to assess the role of the interaction between  ML dipole moments. 
From the values of the total dipole density $D_{\text{tot}}$ given in Table~\ref{table_thermal}, it is apparent that the permanent dipole moments of the two monolayers add up for CAS-1 and CAS-4, whereas the two opposite dipole moments of the monolayers cancel each other for CAS-2 and CAS-3. Each of the monolayers has a dipole density of $\sim 0.025 e$/{\AA}. 
If we assume that the charges contributing to the dipoles are located at the Se and Te atoms of the same monolayer (which are on average 3.4{\AA} apart),  $D_{\text{tot}}$ corresponds to a charge density difference of $q = 0.0073e$/\AA$^2$, or $\pm 0.073e$ per chalcogenide ion. 
Table~\ref{table_thermal} shows the bilayer binding energies calculated for each atomic sequence and stacking pattern under two different treatments of vdW interactions, the pair-wise additive Tkachenko-Scheffler (TS) and the many-body dispersion (MBD) methods. It is seen that for AA-M, AA-X, and AB stacking the interface energy differences are just a few meV, whereas the AA, AB-X, and AB-M bilayers appear to be less stable among the stacking patterns studied in this work. This trend is captured by both TS and MBD methods. Also, among the possible stacking patterns, the AB stacking (two monolayers pointing to opposite directions and atoms with the same x and y registry) has the most negative $E_b$, i.e. the strongest binding for almost all atomic sequences, see Table~\ref{table_thermal}. 
Interestingly, when employing the TS treatment of dispersion forces 
the energetic ordering comes out less clear-cut; AA-X and AA-M become more stable (in CAS-3) or very close in energy to AB (in CAS-4). Also, our TS results suggest that the Se-Se junction  (CAS-3) is thermodynamically the most stable junction among the four possible junctions, but such stability is not reproduced at the MBD level of dispersion-force treatment.

\begin{table*}[tbp]
\centering
\begin{tabular}{SSSSSSSS} \toprule
   {Structure} & {Parameter} & {AA} & {AA-X} & {AA-M} & {AB} & {AB-X} & {AB-M} 
    \\ \midrule\midrule
{} & {$E^{\text{MBD}}_\text{b}~(\text{eV}$)} & {$-0.090$} & {$-0.154$} & {$-0.160$} & {$-0.167$} & {$-0.135$} & {$-0.095$} \\
{CAS-1} & {$E^{\text{TS}}_\text{b}~(\text{eV}$)} & {$-0.11$} & {$-0.295$}& {$-0.301$} & {$-0.307$} & {$-0.277$} & {$-0.215$}  \\
{SeMo{\bf TeSe}WTe} & {$d_\text{inter}~(\text{\AA}$) ($\%$)} & {$4.08~(+4.9$)} & {$3.40~(-0.1)$} & {$3.32~(+0.2)$} & {$3.32~(+0.2)$} & {$3.53 ~(0.0)$} & {$4.04~ (+4.3)$} \\
  {} & {$d^{\text{sub}}_\text{inter}~(\text{\AA}$)} & {$1.82$} & {$1.14$}   & {$1.06$}  & {$1.06$}  & {$1.27$}  & {$1.78$} \\
  {} & {$D_\text{tot}~(10^{-2}e/\text{\AA}$)}        & {$-5.07$} & {$-5.22$} & {$-4.87$} & {$-5.10$} & {$-4.96$} & {$-5.09$}  
\\ \midrule
{} & {$E^{\text{MBD}}_\text{b}~(\text{eV}$)} & {$-0.087$} & {$-0.153$} & {$-0.151$} & {$-0.173$} & {$-0.125$} & {$-0.092$} \\
{CAS-2} & {$E^{\text{TS}}_\text{b}~(\text{eV}$)} & {$-0.210$} & {$-0.298$} & {$-0.296$} & {$-0.315$} & {$-0.271$} & {$-0.215$}  \\
{SeMo{\bf TeTe}WSe} & {$d_\text{inter}~(\text{\AA}$) ($\%$)} & {$4.31~ (+4.0)$} & {$3.62~ (-0.7)$} & {$3.62~ (-0.8)$} & {$3.48 ~(-1.4)$} & {$3.83 ~(+1.1)$} & {$4.26 ~(+3.8)$}\\
  {} & {$d^{\text{sub}}_\text{inter}~(\text{\AA}$)} & {$1.85$} & {$1.16$} & {$1.16$} & {$1.02$} & {$1.37$} & {$1.80$} \\
{} & {$D_\text{tot}~(10^{-2}e/\text{\AA}$)} & {$-0.19$} & {$-0.34$} & {$-0.15$} & {$-0.26$} & {$-0.23$} & {$-0.20$}  
\\ \midrule
{} & {$E^{\text{MBD}}_\text{b}~(\text{eV}$)} & {$-0.090$} & {$-0.161$} & {$-0.159$} & {$-0.161$} & {$-0.144$} & {$-0.095$} \\
{CAS-3} & {$E^{\text{TS}}_\text{b}~(\text{eV}$)} & {$-0.212$} & {$-0.310$} & {$-0.308$} & {$-0.304$} & {$-0.293$} & {$-0.216$}  \\
{TeMo{\bf SeSe}WTe} & {$d_\text{inter}~(\text{\AA}$) ($\%$)} & {$3.88~ (+5.3)$} & {$3.16~ (+5.3)$} & {$3.17~ (+5.1)$} & {$3.21~ (+3.4)$} & {$3.24~ (+6.0)$} & {$3.85~ (+5.4)$}\\
  {} & {$d^{\text{sub}}_\text{inter}~(\text{\AA}$)} & {$1.82$} & {$1.10$} & {$1.11$} & {$1.15$} & {$1.18$} & {$1.79$} \\
{} & {$D_\text{tot}~(10^{-2}e/\text{\AA}$)} & {$0.13$} & {$-0.17$} & {$0.27$} & {$0.06$} & {$0.05$} & {$0.13$}  
\\ \midrule
{} & {$E^{\text{MBD}}_\text{b}~(\text{eV}$)} & {$-0.070$} & {$-0.166$} & {$-0.157$} & {$-0.172$} & {$-0.139$} & {$-0.096$} \\
{CAS-4} & {$E^{\text{TS}}_\text{b}~(\text{eV}$)} & {$-0.150$} & {$-0.312$} & {$-0.302$} & {$-0.317$} & {$-0.284$} & {$-0.220$}  \\
{TeMo{\bf SeTe}WSe} & {$d_\text{inter}~(\text{\AA}$) ($\%$)} & {$4.06~ (+4.6)$} & {$3.30~ (+2.4)$} & {$3.36~ (+0.1)$} & {$3.30~ (+1.4)$} & {$3.49~ (+0.0)$} & {$4.01~ (+4.5)$} \\
  {} & {$d^{\text{sub}}_\text{inter}~(\text{\AA}$)} & {$1.80$} & {$1.04$} & {$1.10$} & {$1.04$} & {$1.23$} & {$1.75$} \\
{} & {$D_\text{tot}~(10^{-2}e/\text{\AA}$)} & {5.10} & {4.63} & {4.99} & {4.85} & {4.73} & {5.01} 
  \\\bottomrule
\end{tabular}
\caption{Thermodynamic stability of Janus MoSeTe/WSeTe bilayer with different chalcogen atom sequences (CAS) and stacking patterns. The chalcogen species forming the junction  inside the bilayer are printed in boldface. The binding energy $E_\text{b}$ (in eV) using different dispersion methods (TS and MBD) are presented followed by the distance $d_\text{inter}$ (in $\text{\AA}$) between the  Janus monolayers of each bilayer structure (for MBD). The deviations (in percent) compared to the TS method are given in parentheses. The distance $d^{\text{sub}}_\text{inter}$ between monolayers after subtraction of the radii of the atoms at the junction and the total electric dipole moment density $D_\text{tot}$ (in $e/\text{\AA}$) of the bilayer stack are shown. }
    \label{table_thermal}
    \end{table*}
 
\par
As pointed out above, the AA and AB-M stacking patterns are the least stable (AA is the least stable and the AB-M the second least stable) and they have the largest inter-layer distance $d_{\text{inter}}$, placing the Janus monolayers further apart. For AA stacking, chalcogen atoms  as well as transition metal atoms of the two Janus monolayers are 
on top of each other, while for AB-M stacking the chalcogens share the same position in the horizontal plane, see Fig.~\ref{fig-sketch}. 
Given the same in-plane positions of atoms, a larger space between the MoSeTe and WSeTe layers is required. 
This is because the lone-pair electronic orbitals of the chalcogens, pointing normal to the plane, repel each other. 
In contrast, for AB stacking, where chalcogen and transition metal atoms of the top ML sit at the hollow site of the bottom ML, more efficient packing of the atoms (less repulsion between filled electronic orbitals) takes place, thus less space is needed. As a result, the inter-layer distance becomes smaller by more than $15~\%$ (relative to that of AA stacking).

\par
Obviously, there is a trivial effect on the junction distance originating from the different sizes of the Se and Te atoms. 
By subtracting the van der Waals radii from the interface distance $d_{\text{inter}}$, one can compensate for this effect and thus obtain insight into the effect of interlayer dipole-dipole interaction on this distance. The van der Waals (atomic) radii of Mo, W, Se, and Te are $190$, $193$, $103$, and $123~\text{pm}$, respectively.  Table~\ref{table_thermal} shows the interlayer distances of the MoSeTe/WSeTe bilayers after  subtraction of these atomic radii  ($d^{\text{sub}}_\text{inter}$). It is seen that, for most stacking patterns (especially AA, AA-M, AA-X, and AB-M), CAS-1 and CAS-4 have the smallest $d^{\text{sub}}_\text{inter}$.  
Conversely, if the same chemical species face each other at the junction 
(CAS-2 and CAS-3), i.e., when the dipole moments of the individual Janus layers point in opposite directions, there is an additional repulsive component to their interaction, and the distance comes out larger by a small amount $\Delta d_{\text{inter}}^{\text{sub}}$. In contrast, when electric dipole moments align (such as in CAS-1 and CAS-4), the MLs attract each other more strongly, 
and the distance is reduced by $\Delta d_{\text{inter}}^{\text{sub}}$ compared to the distance averaged over all four chalcogen atom sequences. 
For the AA, AA-M and AB-M stacking pattern, we can quantify this effect to be approximately $\Delta d_{\text{inter}}^{\text{sub}} = \pm 0.03${\AA}. 
This order of magnitude seems plausible from a simple estimate: The force density acting between the two monolayers has a large elastic contribution plus a smaller contribution from the interaction of the dipole densities. The elastic contribution $f_{\text{elast}} = C \, \Delta d$ can be estimated from the frequency $\omega_{\text{LB}} \approx 20$cm$^{-1}$  of the layer breathing mode (see Sect.~\ref{sec:phonon}) via the relation $C= \rho \omega_{\text{LB}}^2$ where $\rho$ is the reduced mass density.
The electrostatic contribution can be modelled as the force density of a capacitor with two parallel charged plates and amounts to $f_{\text{elec}} = \frac{1}{2} q^2/\epsilon$ where we take $\epsilon$ to be the electrical permittivity of the vacuum. The electrostatic dipole force between the layers displaces them by $\Delta d = \pm f_{\text{elec}}/C = \pm 0.065${\AA} from their equilibrium distance. This value is of the same order of magnitude as $\Delta d_{\text{inter}}^{\text{sub}} = \pm 0.03${\AA} estimated from the structures relaxed with DFT plus MBD vdW treatment. Given that the model is very crude, the level of agreement supports the proposed explanation that electrostatic dipole-dipole interactions are responsible for the changes in interlayer distance.  
\par
We have shown that the stability trends may change by using different levels of dispersion interactions. 
Using the MBD treatment of vdW interactions enables us to more precisely account for the interplay between the self-consistently obtained charge distribution and the fluctuating vdW dipoles. 
As seen from Table~\ref{table_thermal}, the inter-layer distances in the MBD treatment are often larger than in the TS method; this becomes particularly clear for the AA and AB-M stacking. 
These observations are in line with the general finding supported by eq.~\ref{eq-corr-fpa} of mixed attractive and repulsive terms in the expansion of the full electric susceptibility.
This is corroborated by the pronounced distance increase upon using the MBD for CAS-3 (Se-Se junction) where the intrinsic electric dipole moment vectors point toward each other. The larger negative charges accumulated at the junction Se atoms 
give rise to an enhanced screening of vdW dipolar interactions, which is accounted for in the MBD method. 
We recall that in the TS treatment, the Se-Se junction for some stackings was found to be the most stable junction; in hindsight, we judge this to be a faulty overestimation of binding energies by the TS method. 
In fact, with the more realistic polarizabilities of the atomic pairs obtained in the MBD treatment physically more accurate dispersion forces can be obtained, and thus the  attraction or repulsion between the MLs is estimated more accurately.

% This figures shows the vibration spectrum (phonon band structure, layer projected PhDOS ) of bilayer CAS-2: AB.
\begin{figure}[htp]
\centering
\includegraphics[width=0.50\textwidth]{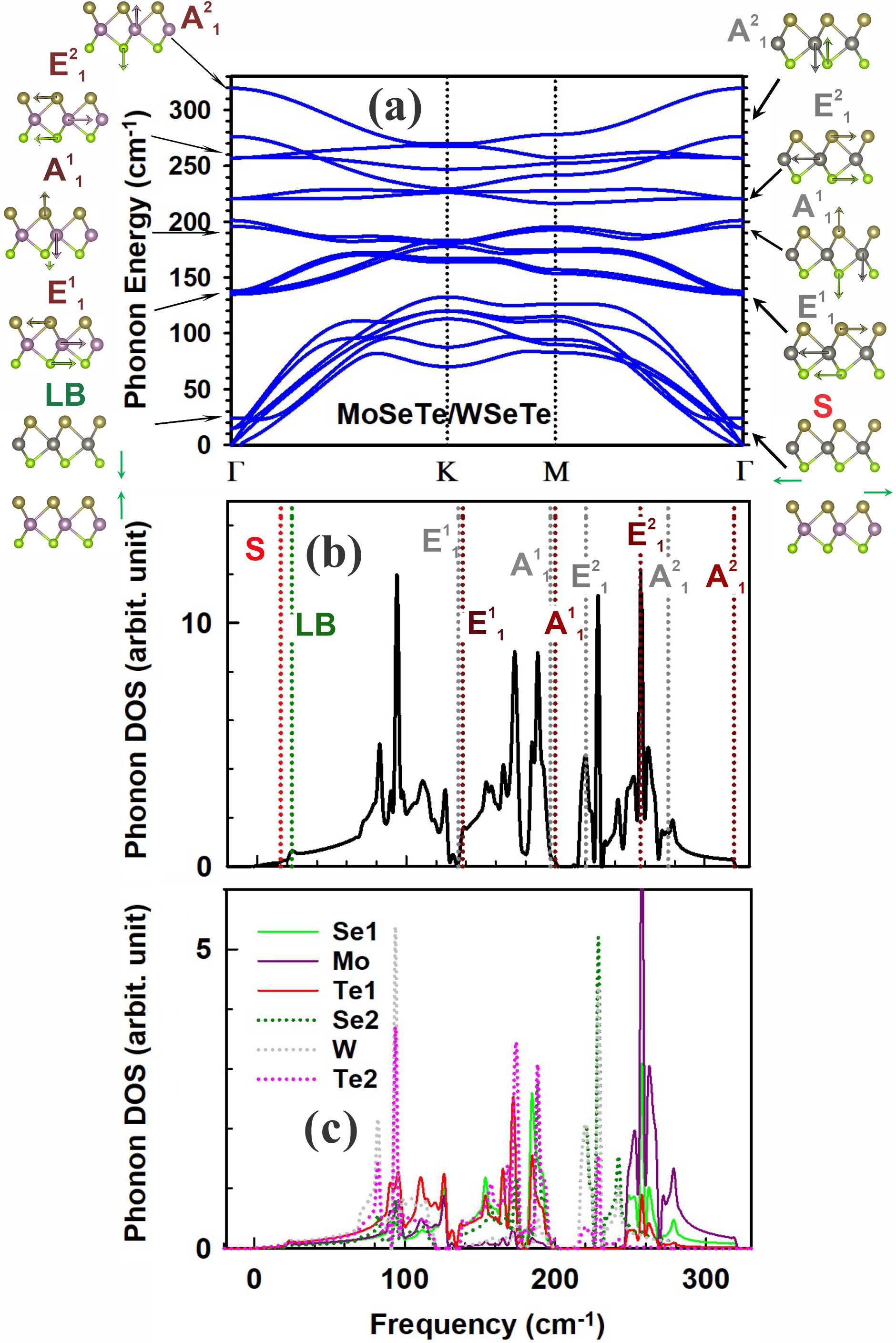}
\caption{Phonon energy bands of the MoSeTe/WSeTe bilayer with AB stacking pattern. The in-plane ($\text{E}^1_1$, $\text{E}^2_1$) and out-of-plane ($\text{A}^1_1$, $\text{A}^2_1$) Raman-active modes and shear (S) and layer-breathing (LB) modes are shown using arrows indicating the directions of atomic displacements. (b) and (c) display the total and atom-projected phonon densities of states, respectively. The major Raman active modes are indicated with vertical dotted lines in (b).}
\label{fig-phoenebands-main-vibrations}
\end{figure}

\subsection{Lattice Vibrations of MoSeTe/WSeTe Bilayers \label{sec:phonon}}

In analogy to other 2D materials, 
it is experimentally feasible to determine  
the thickness (number of layers) of a Janus multi-layer, and to identify the type of junction formed, as well as the stacking pattern, using vibrational spectroscopy~\cite{acsnano-puretzky-15,apl-holler-20,apl-ufuk-22}. 
Therefore theoretical predictions of the vibration characteristics of the Janus multi-layer are useful for the structural characterization of samples. Here, we study the effects of junction (chalcogen atom order) and stacking on the vibration spectrum of Janus MoSeTe/WSeTe bilayers and the changes 
of individual Janus monolayer vibrations upon bilayer formation. 
Moreover, low-lying modes that can be interpreted as relative motions of rigid layers with respect to each other deserve special attention. 

\par
We use the software {\sc Phonopy}\cite{phonopy} as a supplementary package to calculate the vibration spectrum of the bilayers with different chalcogen atom sequences and stacking patterns. After carefully examining the convergence of the $q$-mesh, a $61\times 61 \times 1$ $q$-mesh was used to calculate the phonon 
total and layer-projected, as well as atom-projected, densities of states for every bilayer variant.

\par
For every MoSeTe/WSeTe bilayer geometry, the phonon energy band dispersion was calculated. 
Ignoring small imaginary frequencies below $i5$~cm$^{-1}$ that sometimes occur as a consequence of numerical inaccuracies, 
one can conclude that the relaxed geometries are dynamically stable. 
Since there are six atoms within each primitive unit cell of the Janus bilayer, one can initially perceive 18 vibration modes, out of which six are low-frequency acoustic modes, and the remaining ones are optical modes attributed to the out-of-phase motions of the chalcogen and transition metal atoms. 
Figure~\ref{fig-phoenebands-main-vibrations} displays the major types of atomic vibrations for one bilayer. Four major modes for each monolayer, E$^1$, $\text{A}^{1}_1$, E$^2$, and $\text{A}^{2}_1$, and E$_2$, are Raman-active optical in-plane (E) and out-of-plane (A) vibrations considered to be the fingerprint of Janus monolayers. 
Their energetic positions are marked in Figure~\ref{fig-phoenebands-main-vibrations} (b) and their numerical values can be found in Table~\ref{table_freqallstacking}. 
Due to C$_{\text{3v}}$ structural symmetry of Janus monolayers, the in-plane modes are doubly degenerate. For the MoSeTe/WSeTe bilayer, there are two extra low-frequency vibrations assigned to the gliding and vertical rigid motions of MLs with respect to each other due to coupling caused by the dipole-dipole interactions. These two modes are the shear (S) and layer-breathing (LB) modes which emerge in the low-frequency range below the acoustic region of the vibration spectrum of multi-layer TMDCs. Since rotational C$_{\text{3v}}$ symmetry is preserved in the MoSeTe/WSeTe bilayer, the S mode is doubly degenerate. Also, the frequency of the S mode is calculated to be lower than that of the LB mode since the 
potential energy surface has greater curvature normal than parallel to the bilayer junction. 
Such frequency difference between these two modes remains in place even when the monolayers are non-polar. For example for a {MoSe}$_2$ bilayer with AA-M stacking, the out-of-plane layer-breathing mode has a larger frequency ($31.8~\text{cm}^{-1}$) than the horizontal shear mode ($20.0~\text{cm}^{-1}$).
In cases where the bilayer distance is very large (or the layers are incommensurate), the S modes have (within numerical accuracy) zero frequency. From an experimental perspective~\cite{apl-holler-20}, the S mode is considered a signature of commensurate bilayer stacking. Hence, its absence in the vibration spectrum indicates clearly the lack of lateral restoring forces and thus a 
less corrugated potential energy surface that can result from a 
weak ML binding or a twist in the stacking pattern. Low-frequency Raman spectroscopy is one option to detect such low-energy vibrations~\cite{apl-holler-20,apl-ufuk-22,acsnano-puretzky-15}. 

\par
Figure~\ref{fig-phoenebands-main-vibrations} (c) displays the vibration densities of states projected onto the atoms. According to the mass effect, 
eigenvectors associated with vibrations of a light atom (heavy atom) emerge at the high-frequency (low-frequency) region of the phonon spectrum, see Fig~\ref{fig-phoenebands-main-vibrations} (c). It is seen that the phonon densities-of-state contribution from the Te atom of MoSeTe ML occurs at higher frequencies ($245-280~\text{cm}^{-1}$), while Te atoms of WSeTe ML vibrate with lower frequencies ($240~\text{cm}^{-1}$). 
It is also noted that the Te atom located at the interface vibrates faster than Te at the surface (CAS-1) since the interface Te atom feels an additional restoring force due to the  inter-layer interaction. 

Next, we turn to the analysis of the low-frequency bilayer modes since these turn out to be most sensitive to inter-layer coupling. 
For AA and AB-M stacking the frequency of the S mode is found to be zero within numerical accuracy, 
see Table~\ref{table_freqallstacking}. This resonates well with thermodynamic stability (see Table~\ref{table_thermal}) analysis pointing to thermodynamically less stable AA and AB-M stacking, 
and also in good agreement with low-energy Raman spectroscopy measurements~\cite{apl-holler-20}, which suggest that the S mode is 
a zero-energy mode 
for AA stacking (R-type stacking) of a hetero-bilayer, while there is a non-zero frequency (above $10~\text{cm}^{-1}$) assigned to the LB mode, see Table~\ref{table_freqallstacking}. 
This indicates that the restoring force of rigid gliding vibration is weaker at the AA stacking, which is a clear manifestation of less strong binding of MLs, see Table~\ref{table_thermal}. Also, for every atomic sequence  except for 
CAS-3, the S mode of AA-M (3R type) occurs at lower frequencies compared to the AB (2H type) S mode. This is similar to the experimental results obtained for the homo-bilayer of MoSe$_2$~\cite{acsnano-puretzky-15}. 

\par
\begin{figure}[htp]
\centering
\includegraphics[width=0.50\textwidth]{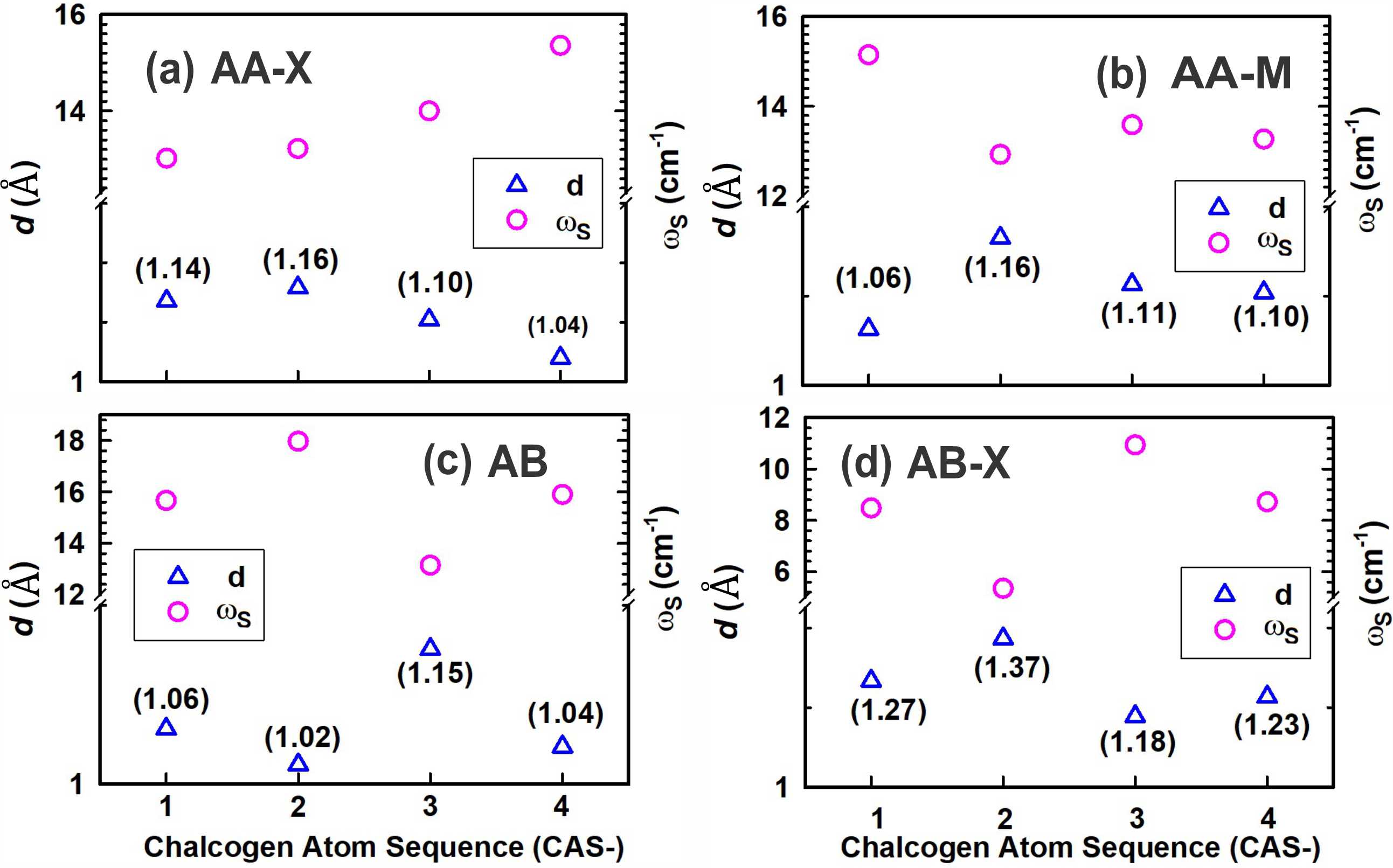}
\caption{Changes in the interlayer distance $d$ (triangles) and shear-mode frequency  $\omega_{\text{S}}$ (circles) of MoSeTe/WSeTe bilayers with different stacking patterns: AA-X (a), AA-M (b), AB (c), and AB-X (d) with varying the chalcogen atom sequence. 
}
\label{fig-vib-s-mode-vs-d-diff-cas}
\end{figure}

The systematic dependence on the chalcogen atom sequence (CAS) is shown in Figure~\ref{fig-vib-s-mode-vs-d-diff-cas} for both the change in inter-layer distance and the energy of the S mode. Only four stacking patterns are included (panels (a) to (d)), since 
for the remaining two stackings, AA and AB-M  
the parallel component of the restoring force is too weak,  as pointed out earlier. 
We want to stress that the 
inter-layer distance $d$ and the S mode frequency $\omega_{\text{S}}$ are anti-correlated when varying the CAS. For CAS-1 and CAS-4 bilayers, both having non-zero net electric dipole moments and thus 
allow for stronger 
attraction, $d$ becomes shorter. Since the MLs are coupled stronger by this  attraction, 
a more corrugated potential and stiffer horizontal rigid motions of the MLs are expected, and thus the S mode frequency increases. This line of argument rationalizes the trends visible in Fig.~\ref{fig-vib-s-mode-vs-d-diff-cas}. 

Finally, we point out that Raman spectroscopy of the optical modes could be used, too, to identify differences between stacking patterns. 
It has been demonstrated that, by using conventional vibrational spectroscopy techniques, frequency shifts as small as $1~\text{cm}^{-1}$ can be measured for Raman-active modes~\cite{acsnano-puretzky-15,apl-holler-20,apl-elis-11}. Looking at the frequencies of the major Raman-active vibration mode A$_1$ of a bilayer, one can see that for all atomic sequence variants,  except for CAS-2 (the Te1-Te2 junction),  the A$_1$ of WSeTe is higher compared to the isolated single layer, see Table~\ref{table_freqallstacking}. 
This can well be rationalized using the classical model for coupled harmonic oscillators (CHO). According to the CHO model, established dipole-dipole interaction between the atoms of the monolayer increases the effective restoring force acting on the atoms, which stiffens the out-of-plane vibrations, and thus the associated frequency increases.
Also, the energy splitting of the out-of-plane A$_1$ modes of Janus MoSeTe and WSeTe monolayers at the $\Gamma$-point tends to become largest for CAS-1 and CAS-4 bilayers and smallest for CAS-2 and CAS-3. This is the case for the AA, AA-M, AB, and AB-X stacking patterns, see Table~\ref{table_freqallstacking}. Such enhancement in the A$_1$ energy difference points to the stronger coupling between the out-of-plane vibrations, which as well confirms the attractive interaction of electric dipole moments aligned in the same direction in both the MoSeTe and the WSeTe layers in CAS-1 and CAS-4 bilayers.

Comparing the vibration spectra obtained using different treatments of dispersion interactions can give us a decent clue which treatment may provide a more realistic description of the interactions between induced charge polarization. Table~\ref{table_freq-ts-vs-mbd} gives the frequencies of the major Raman-active vibration modes obtained using the TS and MBD treatments of polarization-induced interactions. For Janus MoSeTe ML, the out-of-plane vibration modes ($\text{A}^1_1$ and $\text{A}^2_1$) shift by $1$ to $2~\text{cm}^{-1}$ to higher frequencies. This can be explained by the fact that under the MBD treatment, the ML is slightly squeezed normal to the ML plane due to a larger attraction between the transition metal and the chalcogenides. In contrast, the in-plane atomic distances become slightly larger. This softens the in-plane vibrations ($\text{E}^1_1$ and $\text{E}^2_1$), and thus lower  frequencies of the in-plane vibrations are expected, see Table~\ref{table_freq-ts-vs-mbd}. However, it is clear that such frequency shifts for the in-plane and out-of-plane vibration modes of Janus WSeTe ML remain small since the pertinent changes in the bond lengths tend to be negligible. 

% This figure shows that changes in the distances and frequencies of shear mode due to changing the atomic sequence

% This table shows the low- and high-frequency Raman modes of bilayers with different atomic sequences and stacking.
\begin{table*}[tbp]
\centering
\begin{tabular}{SSSSS} \toprule
 {Atomic sequence} & {Stacking} & {High-frequency Raman modes  $(\text{cm}^{-1})$ } & \multicolumn{2}{c}{Low-frequency  $(\text{cm}^{-1})$ } \\
   {}  &   {} & { $E^{1}_1$, $A^{1}_1$, $E^{2}_1$, $A^{2}_1$} & S & LB \\
   \midrule\midrule
{Isolated MoSeTe} & {} & {$134.8$, \hspace{1.00cm} $197.1$, \hspace{1.00cm} $257.1$, \hspace{1.00cm} $321.3$ \hspace{0.9cm} } & {n.a.} & {n.a.} \\
{Isolated WSeTe} & {}  & {\hspace{0.50cm} $(137.1)$, \hspace{0.8cm} $(198.9)$, \hspace{0.80cm} $(221.1)$, \hspace{0.80cm} $(277.7)$}  & {n.a.} & {n.a.}  \\
{} & {AA} & {$134.7~(137.1)$, $196.3~(199.3)$, $256.9 ~(221.0)$, $319.2~ (277.1)$}   & 0 &    18.3  \\

{} &{AA-X}   & {$134.9~ (137.8)$, $198.0~ (198.0)$, $257.1~ (220.7)$, $319.6~ (276.2)$}   & 13.0 & 21.5  \\

{CAS-1}    & {AA-M}   & {$135.0~ (137.8)$, $196.0 ~(201.2)$, $257.1~ (220.7)$, $319.6~ (276.4)$}   & 15.2 & 23.9  \\

{SeMo{\bf TeSe}WTe} & {AB}     & {$135.8~ (137.8)$, $196.6~ (199.8)$, $257.0~ (220.4)$, $319.7~ (275.8)$}   &  15.7 & 23.2   \\

{}         & {AB-X}   & {$134.9~ (137.2)$, $196.9 ~(198.8)$, $257.0~ (220.7)$, $319.5~ (278.3)$}   & 8.5 & 19.7  \\

{}         & {AB-M}   & {$134.7~ (137.2)$, $196.3~ (199.1)$, $257.2~ (221.2)$, $319.8~ (277.8)$} & 0 & 17.4    \\ 
\midrule
{}         & {AA}     & {$134.7~ (137.0)$, $196.2 ~(198.4)$, $257.0~ (221.2)$, $319.4~ (277.8)$} &  0 &  17.9  \\

{}         & {AA-X}   & {$134.9~ (137.3)$, $196.1~ (197.7)$, $257.0~ (221.0)$, $320.1~ (277.4)$} & 13.2 & 23.4  \\

{CAS-2}         & {AA-M}   & {$134.8~ (137.1)$, $196.1~ (198.8)$, $257.0~ (221.0)$, $319.3~ (278.1)$} &  12.9 & 19.0  \\

{SeMo{\bf TeTe}WSe}         & {AB}     & {$134.8 ~(137.1)$, $196.0~ (197.8)$, $257.0~ (221.0)$, $319.6~ (277.4)$} &  18.0 & 23.8  \\

{}         & {AB-X}   & {$134.7~ (137.1)$, $196.1~ (198.7)$, $257.0~ (221.1)$, $320.1~ (278.0)$} &  5.4 & 17.3  \\

{}         & {AB-M}   & {$134.7~ (137.0)$, $195.1~ (198.1)$, $257.2~ (221.3)$, $319.8~ (277.9)$} &  0 & 22.4  \\ 
\midrule
{}         & {AA}     & {$134.7~ (137.1)$, $196.7~ (199.2)$, $256.8~ (220.9)$, $319.7~ (277.2)$} &  0 & 17.9  \\

{}         & {AA-X}   & {$135.1~ (137.9)$, $196.4~ (201.1)$, $256.6~ (221.0)$, $319.3~ (276.7)$} & 14.0 & 19.1  \\

{CAS-3}         & {AA-M}   & {$135.5~ (137.6)$, $197.0~ (199.6)$, $256.8~ (221.2)$, $318.5~ (277.6)$} &  13.6 & 33.6  \\

{TeMo{\bf SeSe}WTe}         & {AB}     & {$135.3~ (137.6)$, $197.4~ (199.1)$, $256.5~ (220.7)$, $318.7 ~(278.6)$} &  13.2 & 30.3  \\

{}         & {AB-X}   & {$135.0~ (137.3)$, $197.0~ (199.2)$, $256.6~ (220.8)$, $319.1~ (277.6)$} &  10.9 &19.0  \\

{}         & {AB-M}   & {$134.8~ (137.1)$, $197.4~ (201.0)$, $257.1~ (221.0)$, $318.5~ (277.6)$} &  0 &18.3  \\
\midrule
{}         & {AA}     & {$135.4~ (137.1)$, $197.0~ (201.4)$, $260.7~ (223.9)$, $324.1~ (282.3)$} &  0 & 24.4  \\

{}         & {AA-X}   & {$135.1~ (137.4)$, $197.0~ (198.7)$, $256.3~ (221.1)$, $318.8~ (277.3)$} &  15.4 & 23.7   \\

{CAS-4}         & {AA-M}   & {$135.5~ (137.4)$, $192.3~ (199.6)$, $256.5~ (221.1)$, $317.2~ (277.9)$} &  13.3 & 24.7  \\

{TeMo{\bf SeTe}WSe}         & {AB}     & {$135.4~ (137.1)$, $197.0~ (198.6)$, $256.2~ (221.1)$, $318.1~ (277.4)$} &  15.9 & 24.1  \\

{}         & {AB-X}   & {$134.9~ (137.1)$, $193.1~ (202.2)$, $256.6~ (221.0)$, $315.6~ (277.7)$}  &  8.7 & 26.2  \\

{}         & {AB-M}   & {$134.9~(137.0)$, $196.9~(198.7)$, $257.2~ (221.2)$, $319.4~ (277.3)$} &  0 &18.6  \\ 
\bottomrule
\end{tabular}
\caption{Frequencies of the Raman-active modes (third column) and of horizontal shearing (S) and vertical layer-breathing (LB) modes  (right column) of MoSeTe/WSeTe bilayers with different chalcogen atom sequences and stacking patterns. The frequencies of the Raman active modes of isolated MoSeTe and WSeTe monolayers are given as a reference in the first two rows. These data were obtained using the MBD treatment of the dispersion forces. Here, the vibration frequencies of WSeTe are given in parentheses.}
    \label{table_freqallstacking}
    \end{table*}
   
% This table shows the frequencies of Raman modes of each layer at different levels of vdW theory; TS, and MBD. 
\begin{table}[tbp]
\centering
\begin{tabular}{llrrrr} \toprule
   {structure} & & \multicolumn{4}{c}{Raman modes  $(\text{cm}^{-1})$} \\
     & & $\text{E}^1_1$ & $\text{A}^1_1$ & $\text{E}^2_1$ & $\text{A}^2_1$  \\ \midrule\midrule
{MoSeTe} & TS  & 136.1 & 195.3 & 259.5 & 320.1 \\
         & MBD & 134.8 & 197.1 & 257.1 & 321.3 \\ \midrule
{WSeTe}  & TS & 138.3 & 198.6 & 223.0 & 278.1 \\
         & MBD & 137.1 & 198.9 & 221.1 & 277.7  \\ \bottomrule

\end{tabular}
\caption{Frequencies of the Raman-active modes for isolated Janus MoSeTe and WSeTe MLs obtained with pair-wise additive Tkatchenko-Scheffler (TS) and many-body-dispersion (MBD) treatments of vdW interaction.}
    \label{table_freq-ts-vs-mbd}
    \end{table}

% This is the figure for comparing the band structures of isolated ML, bilayer w/o vdw interaction and with vdw interaction.
\begin{figure}[ht]
\centering
\includegraphics[width=0.50\textwidth]{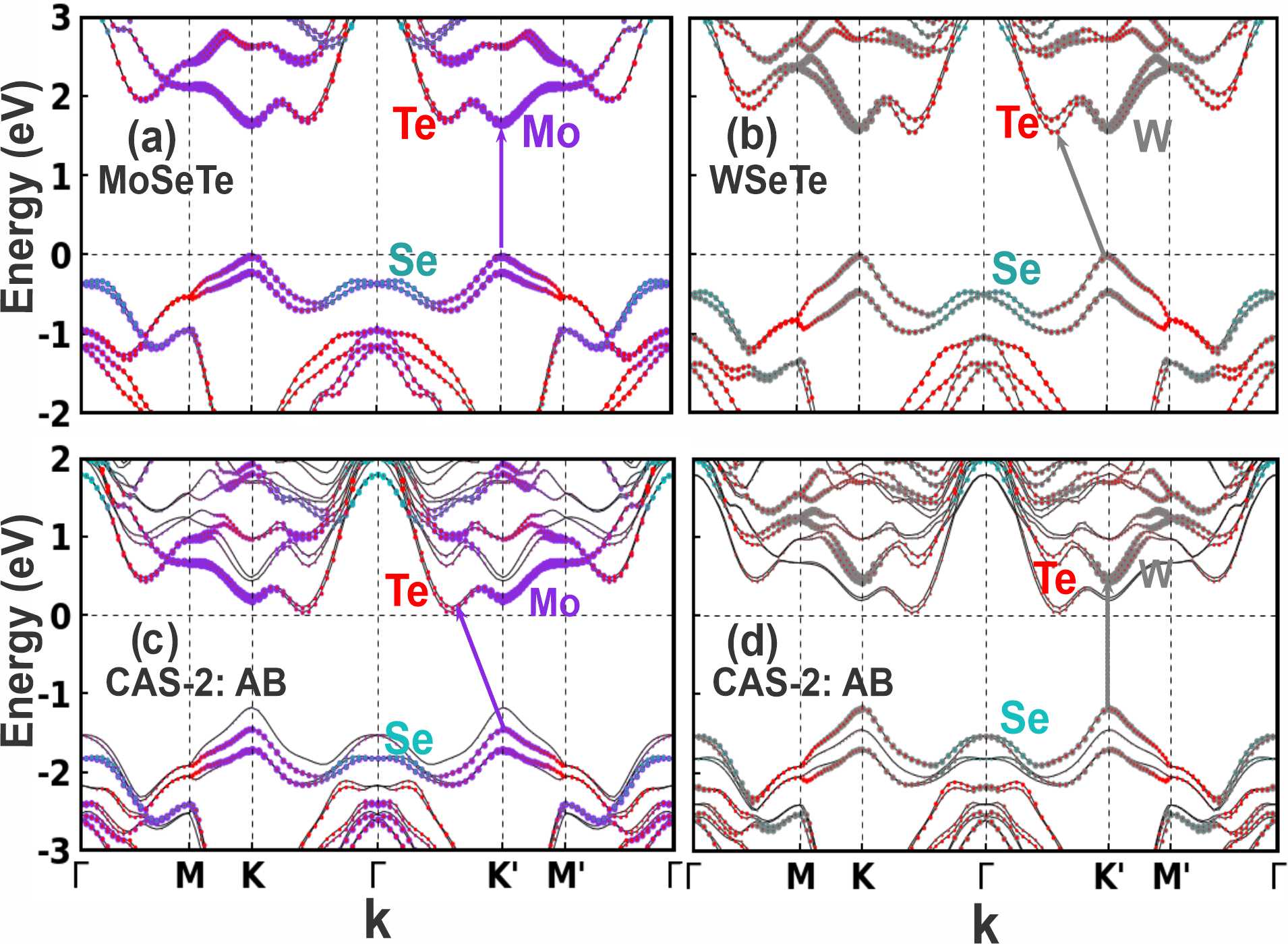}
\caption{Electronic band structures of isolated MoSeTe (a) and WSeTe (b) monolayers, and a typical MoSeTe/WSeTe bilayer projected onto MoSeTe (c) and WSeTe (d) monolayers and their atomic constituents. The contribution from Mo, W, Se, and Te atoms are displayed in purple, gray, green, and red colors, respectively. Here, the Fermi energy is set to zero. Also, the color arrows indicate the type of band gap within each monolayer. The van der Waals interactions are treated using the MBD method.}
\label{fig-ele-bands}
\end{figure}

\subsection{Electronic structure of MoSeTe/WSeTe bilayers with different stacking}

As we already explained, two Janus MLs can only be held together by the van der Waals interaction, since the surface of each Janus ML is terminated by fully occupied p$_z$ orbitals, i.e. by electron lone pairs. 
The van der Waals interactions are non-covalent; thus one expects that the electronic band structure of a MoSeTe/WSeTe bilayer is essentially a superposition of the band structures of the isolated MoSeTe and WSeTe MLs; however, with an unknown energetic offset.  
To illustrate this point, we discuss one example, the MoSeTe/WSeTe bilayer (CAS-2: AB). In Fig.~\ref{fig-ele-bands}, we show first the band structures of the two monolayers MoSeTe and WSeTe on their own, 
followed by the MoSeTe/WSeTe band structure at the optimized interlayer distance using the MBD method for the vdW interactions. 
Looking closer at the orbital projections of the bands, we learn that a type-II band alignment is realized, i.e., the valence band maximum (VBM) and conduction band minimum (CBM) are formed by orbitals from different Janus MLs. 
However, as we will show in the following, there are subtle changes in the band structure depending on the chalcogen atom sequence. 
For example, for an isolated Janus MoSeTe monolayer, both the VBM and the CBM appear at the high-symmetry point $K$. This makes the Janus MoSeTe monolayer a direct band gap semiconductor. Similar to the well-known case of MoS$_2$~\cite{Fang2015}, also in MoSeTe the VBM and CBM are mainly formed by the Mo d$_{x^2 -y^2}$ and d$_{z^2}$ orbitals, respectively. 
The property of the band gap (being direct) may change when the Janus MoSeTe and WSeTe monolayers form a bilayer,  
as the minimum of the conduction band between $\Gamma$ and $K$ is now somewhat lower than the $K$ point. This is a consequence of the p-orbitals of the Te atoms from the lower and upper layers overlapping, thus giving stronger dispersion to the conduction band. 
Contrarily, in the MoSeTe monolayer, one could say that the higher degree of quantum confinement reduces the band dispersion; therefore the monolayer displays a direct band gap.

\begin{table*}[tbp]
\centering
\begin{tabular}{SSSSSSSS} \toprule
   {Structure} & {Quantity} & {AA} & {AA-X} & {AA-M} & {AB} & {AB-X} & {AB-M} 
    \\ \midrule\midrule
{CAS-1} & {$E_\text{g}~(\text{eV})$} & {1.19} & {1.11} & {1.24} & {1.15} & {1.21} & {1.18}   \\
{SeMo{ \bf TeSe}WTe} & {$\Delta E_\text{CBM}~(\text{eV})$} & {0.312}  & {0.354} & {0.271} & {0.334} & {0.287} & {0.322} \\\midrule

{CAS-2} & {$E_\text{g}~(\text{eV})$} & {1.28} & {1.32} & {1.25} & {1.24} & {1.29} & {1.31}  \\
{SeMo{\bf TeTe}WSe} & {$\Delta E_\text{CBM}~(\text{eV})$} & {0.193}  & {0.167} & {0.197} & {0.171} & {0.186} & {0.194} \\\midrule

{CAS-3} & {$E_\text{g}~(\text{eV})$} & {\underline{1.32}} & {1.16} & {1.12} & {1.19} & {1.16} & {\underline{1.33}}  \\
{TeMo{\bf SeSe}WTe} & {$\Delta E_\text{CBM}~(\text{eV})$} & {0.217}  & {0.110} & {0.271} & {0.191} & {0.188} & {0.220} \\\midrule

{CAS-4} & {$E_\text{g}~(\text{eV})$} & {0.60} & {\underline{0.73}} & {\underline{0.62}} & {\underline{0.66}} & {\underline{0.69}} & {\underline{0.63}}  \\
{TeMo{\bf SeTe}WSe} & {$\Delta E_\text{CBM}~(\text{eV})$} & {0.873}  & {0.703} & {0.781} & {0.816} & {0.799} & {0.837}  \\\bottomrule
\end{tabular}
\caption{Electronic properties of Janus MoSeTe/WSeTe bilayer with different chalcogen atom sequences (CAS) and stacking patterns. The chalcogen species forming the interface inside the bilayer are printed in boldface. 
The band gap $E_\text{g}$ (in eV), and the offset of the conduction band minimum $\Delta E_\text{CBM}$ (in eV) at the high-symmetry point $K$ are listed. The values for $E_\text{g}$ are underlined for direct band gaps.}
    \label{table_electronics}
    \end{table*}

\par
\begin{figure}[ht]
\centering
\includegraphics[width=0.350\textwidth]{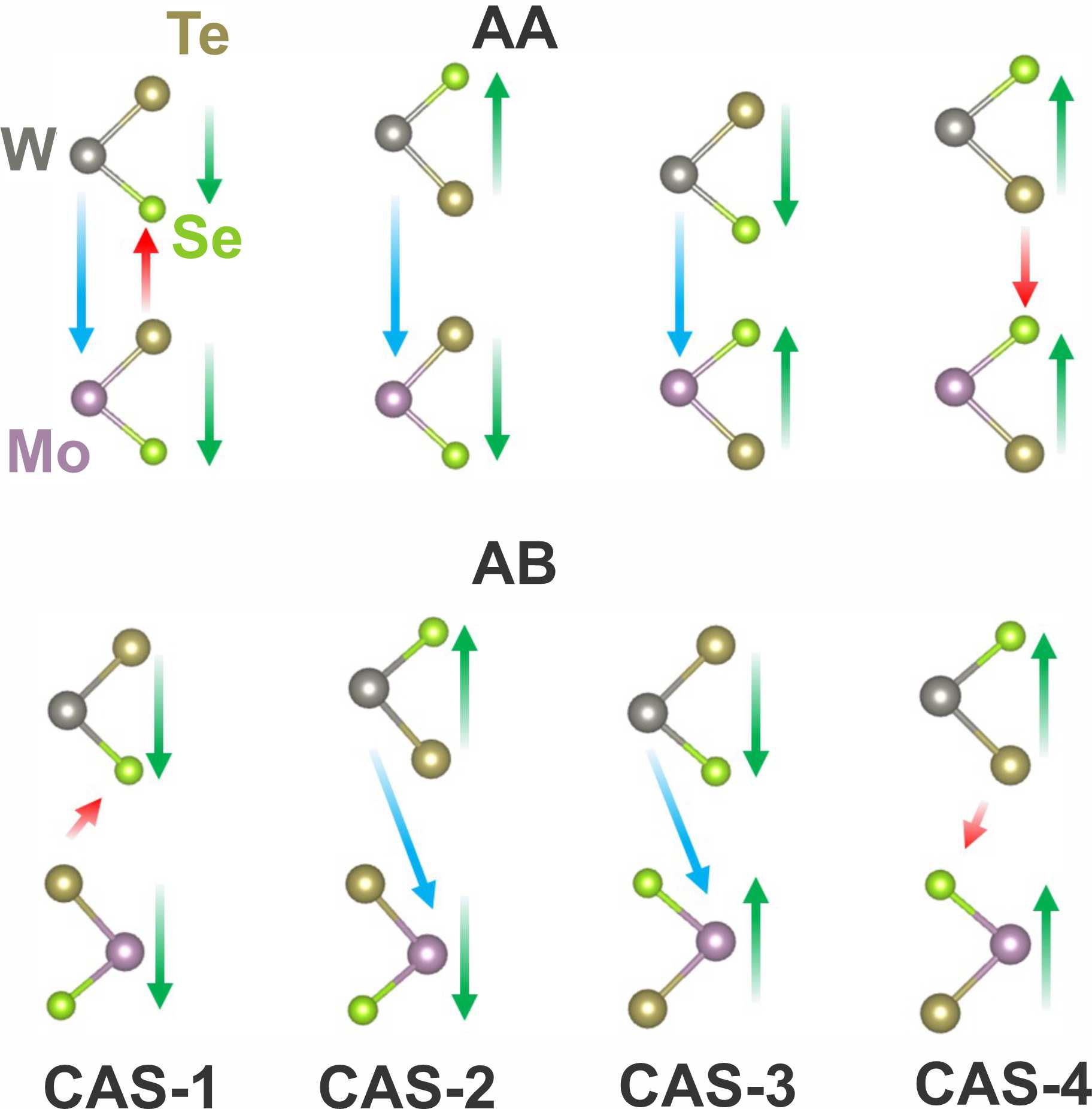}
\caption{Electronic dipole moments for the chalcogen atomic sequences of a MoSeTe/WSeTe bilayer with AA (top row) and AB stacking (bottom row).  The color code for the atomic species is the same as in Fig.~\ref{fig-sketch}.}
\label{fig-dipole-aa-ab-stackings}
\end{figure}

Next, we turn to a general discussion of the electronic structure of the bilayers for all possible chalcogen atom sequences and stackings. An overview of the data is given in Tab.~\ref{table_electronics}. Table~\ref{table_thermal}  contains the total  areal electronic dipole densities of the MoSeTe/WSeTe bilayers studied in this work. 
Since the isolated Janus MoSeTe and WSeTe monolayers lack out-of-plane inversion symmetry and thus are polar, they possess already dipole moments densities of $-2.51$ and $-2.40 \times 10^{-2}\text{e}/\text{\AA}$, respectively. If the interface inside the bilayer consists of different atomic species Se and Te, as for CAS-1 and CAS-4, the intrinsic dipole vectors of the MLs point in the same direction, adding up to nearly the sum of individual dipole moments, 
as illustrated by the green arrows in Fig.~\ref{fig-dipole-aa-ab-stackings}. 
In contrast, when the junction is formed by atoms of the same type, the total dipole density becomes very small ($< 3 \times 10^{-3}~\text{e}/\text{\AA}$). 
Most notably, the energy gap of the heterostructure bilayers varies strongly with the CAS, being largest for CAS-2 and CAS-3, having the same species at both sides of the junction, somewhat smaller for CAS-1 and smallest ($\sim 0.6~\text{eV}$) for CAS-4. These findings can be rationalized by considering them in the context of the dipole moment densities of the MoSeTe and WSeTe monolayers: 
If the dipoles point from W to Mo, as in CAS-1, the overall higher-lying conduction band of WSeTe is pushed down below the Mo-derived conduction band, resulting in a larger band gap of the bilayer heterostructure, at least in AA stacking. 
However, if the dipole moments are pointing from Mo to W, as in CAS-4, the Mo-derived conduction band is shifted down due to electrostatic effects, leading to a strongly reduced band gap of the bilayer heterostructure.  
These considerations are still somewhat simplified. 
To fully account for the variation of the dipole moment densities not only with chalcogen atom sequence but also with the various stacking patterns, one needs to consider the additional dipole moments that arise when the two MLs are brought into contact (blue and red arrows in Fig.~\ref{fig-dipole-aa-ab-stackings}). These  additional dipole moments are not necessarily pointing perpendicular to the junction and both may strengthen or weaken each other, thus giving rise to a more complex dependence on stacking patterns.

% This is the figure that displays the dipole moments components contributing to the total dipole moments of bilayer

\begin{figure*}[tb]
\centering
\includegraphics[width=1.000\textwidth]{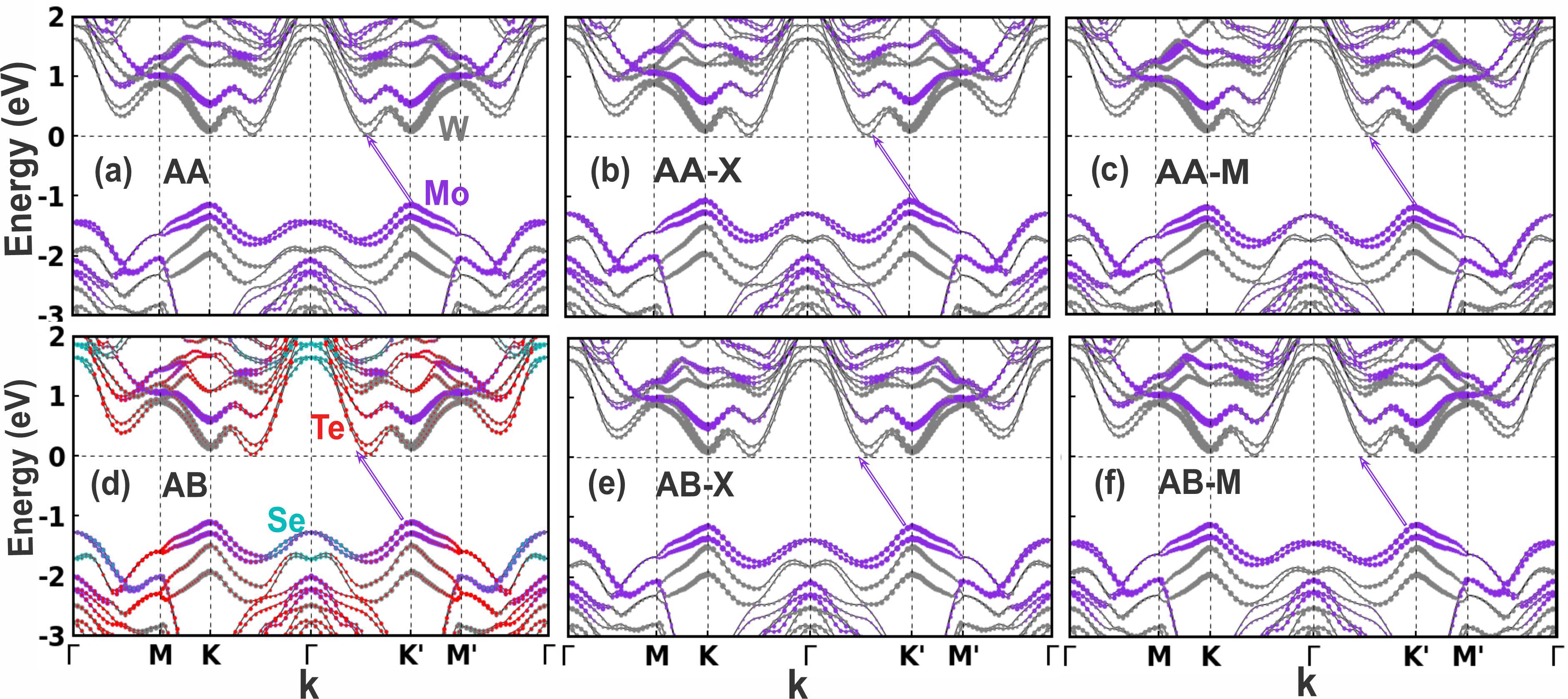}
\caption{Electron energy band structure of MoSeTe/WSeTe bilayer with chalcogen atom sequence CAS-1 and different stackings. Here, the purple, gray, cyan, and red colors reflect the orbital contributions from Mo, W, Se, and Te atoms, respectively. Here, the Fermi energy is set to zero.}
\label{fig-bands-cas-1}
\end{figure*}

\begin{figure*}[tb]
\centering
\includegraphics[width=1.000\textwidth]{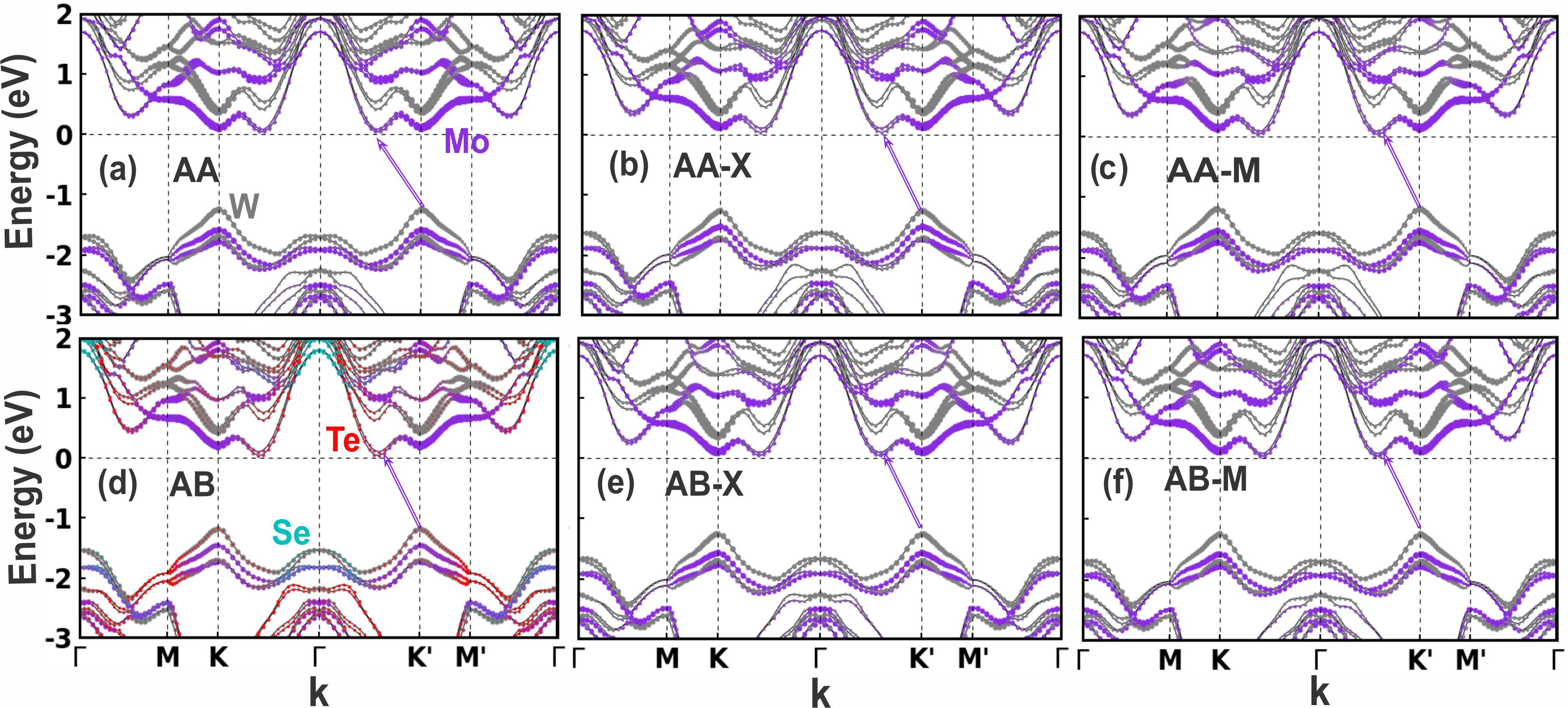}
\caption{Same plots as in Fig.~\ref{fig-bands-cas-1}, but for chalcogen atom sequence CAS-2.}
\label{fig-bands-cas-2}
\end{figure*}

\begin{figure*}[tb]
\centering
\includegraphics[width=1.000\textwidth]{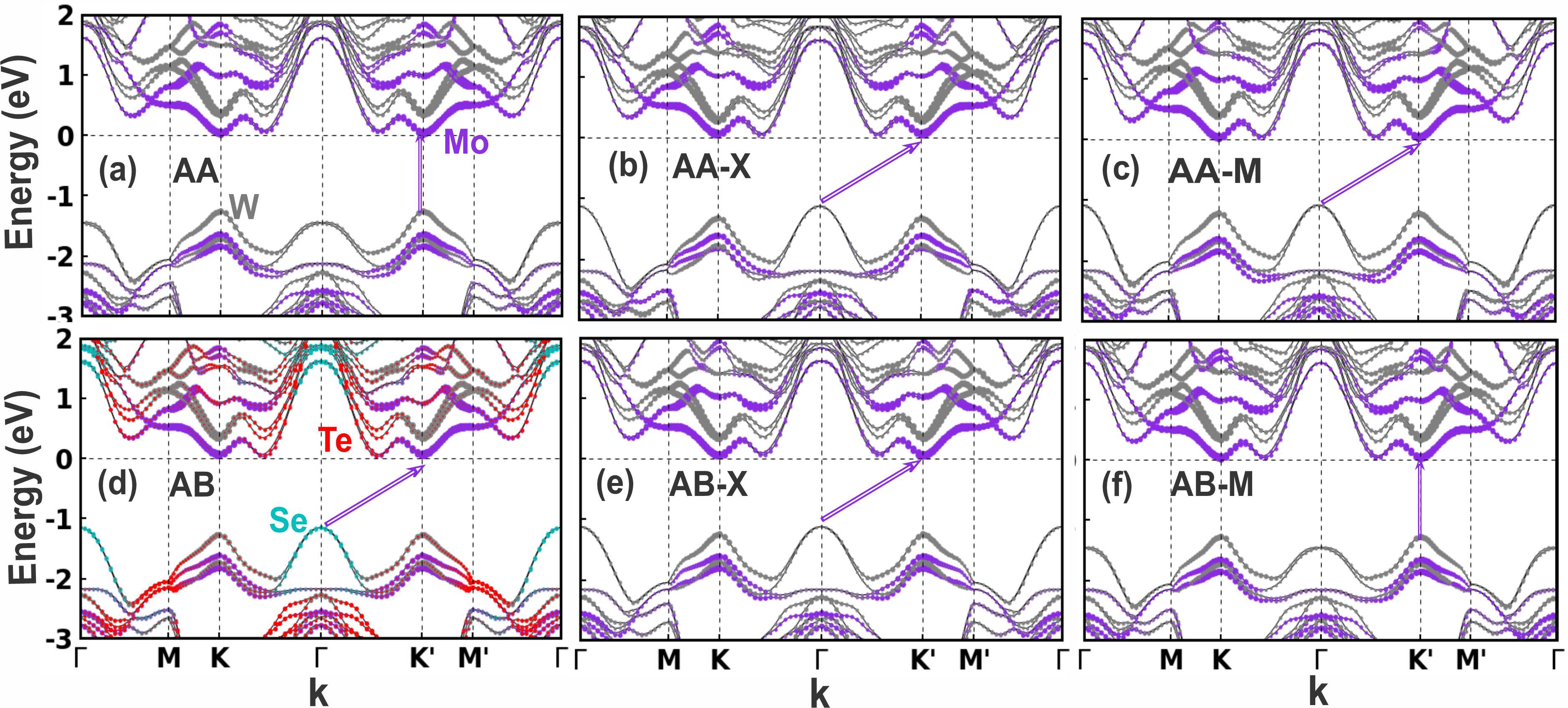}
\caption{Same plots as in Fig.~\ref{fig-bands-cas-1}, but for chalcogen atom sequence CAS-3.}
\label{fig-bands-cas-3}
\end{figure*}

\begin{figure*}[tb]
\centering
\includegraphics[width=1.000\textwidth]{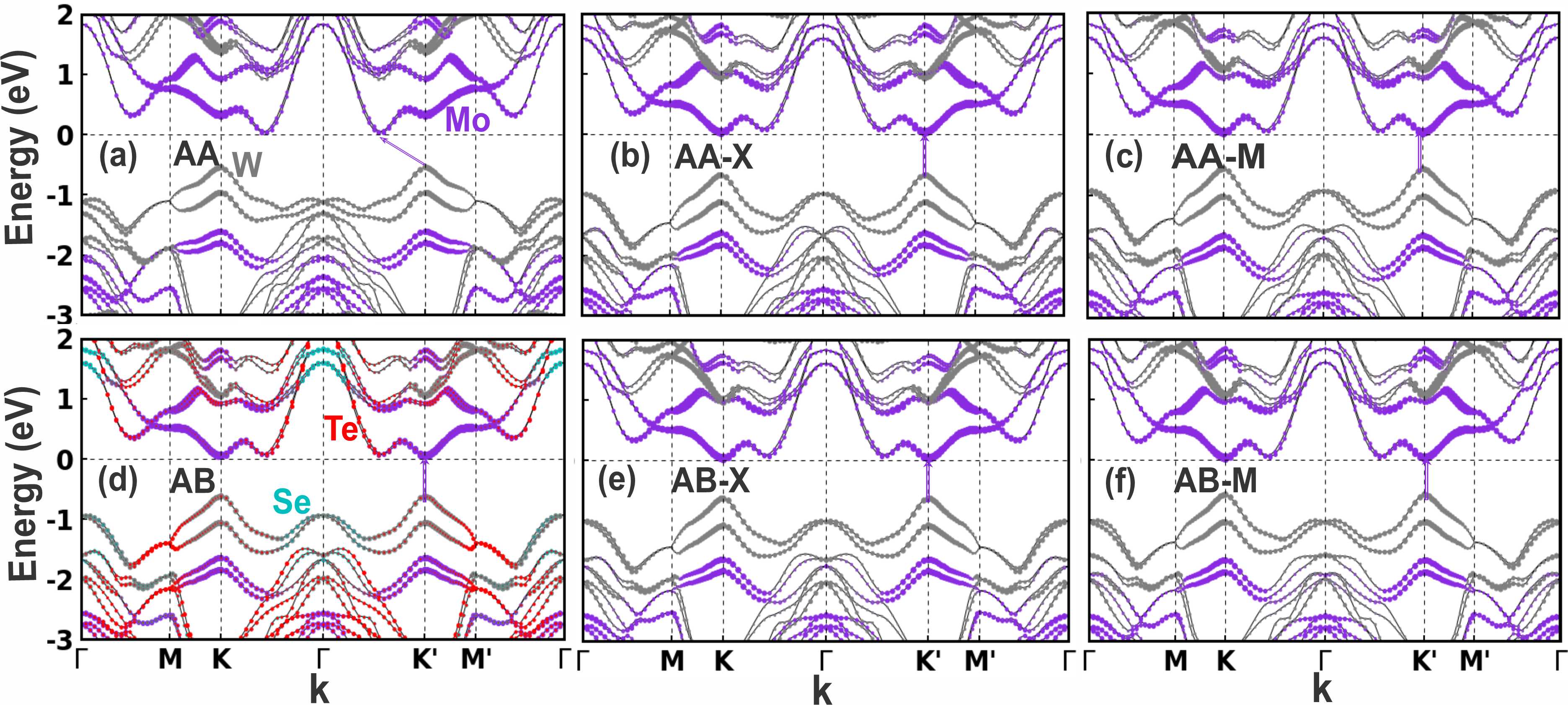}
\caption{Same plots as in Fig.~\ref{fig-bands-cas-1}, but for chalcogen atom sequence CAS-4.}
\label{fig-bands-cas-4}
\end{figure*}

Deeper insight into the band alignment in the bilayer can be obtained by looking at the band structures. 
Figures~\ref{fig-bands-cas-1} to \ref{fig-bands-cas-4} display the electronic band structures along a high-symmetry $k$-path ($\Gamma \rightarrow M \rightarrow K \rightarrow \Gamma \rightarrow K' \rightarrow M' \rightarrow \Gamma$) for the MoSeTe/WSeTe bilayers of different chalcogen atom sequences and stacking patterns.  
Moreover, by projecting the bands onto atomic orbitals it is possible to identify which band originates from which layer in the heterostructure. 
Looking at the energy bands of isolated Janus MLs, see Fig.~\ref{fig-ele-bands}, we noted already that the minimum conduction and maximum valence bands, with the exception of  the $\Gamma$ point (where p orbitals of Se atoms form the bands), are formed mostly by d and p electronic orbitals of transition metal and Te atoms, respectively. 
MoSeTe has a direct band gap at the K point, but for the WSeTe monolayer, the CBM emerges at a point $Q$ within the $K\rightarrow\Gamma$ path. This is consistent with the results of scanning tunneling microscopy of a WSe$_2$ monolayer which suggests that the CB at the point $Q$ (between the $K$ and $\Gamma$ points) stays a bit lower in energy relative to the K point, and thus, somewhat  surprisingly, the WSe$_2$ possesses an indirect band gap~\cite{nanolett-zhang-15},  in contrast to most other TMDC monolayers.

% This Table shows energy band splitting of valence band at the high-symmetry point K $\Delta E^{\text{SOC}}_V$ for Mo and W atoms.
\begin{table*}[tbp]
\centering
\begin{tabular}{SSSSSSS} \toprule
   {structure} & {AA}  & {AA-X} & {AA-M} & {AB} & {AB-X} & {AB-M} 
    \\ \midrule\midrule
{CAS-1} & {0.200 (0.453)}  & {0.201 (0.456)} & {0.193 (0.449)} & {0.178 (0.435)} & {0.200 (0.455)} & {0.199 (0.453)} \\
{CAS-2} & {0.205 (0.445)}  & {0.205 (0.454)} & {0.199 (0.455)} & {0.250 (0.498)} & {0.202 (0.452)} & {0.199 (0.451)} \\
{CAS-3} & {0.200 (0.455)}  & {0.202 (0.458)} & {0.200 (0.455)} & {0.207 (0.461)} & {0.201 (0.456)} & {0.200 (0.454)}\\
{CAS-4} & {0.193 (0.415)}  & {0.199 (0.452)} & {0.200 (0.454)} & {0.197 (0.435)} & {0.200 (0.456)} & {0.200 (0.452)} \\ \bottomrule
\end{tabular}
\caption{Valence band splitting $\Delta E^{\text{SOC}}_V$ at the high-symmetry point $K$ for Mo and W atoms. The numbers for W are given in parentheses.}
    \label{table_electronics_splitting}
    \end{table*}

Even when a bilayer is formed, the optical matrix elements for transitions between occupied and unoccupied bands are still much larger for the bands originating from the {\em same}  monolayer, i.e. for intra-layer transitions, compared to inter-layer transitions. Hence, the question arises if the optical transitions in a hetero-bilayer follow the same rules as for a monolayer, or if significant changes in the band structure of one layer arise due to the proximity of the other layer. 
This type of behavior is well known for bilayers of MoS$_2$: In contrast to a MoS$_2$ monolayer,  the bilayer possesses an indirect gap, as demonstrated by 
experiments using absorption, photoluminescence, or photoconductivity measurements~\cite{prl-mak-10, prv-hand-11}, as well as by electronic structure calculations~\cite{apl-elis-11}. Likewise, we find that the electronic band gap of a bilayer MoSeTe/WSeTe becomes indirect from 
$K$ to $Q$, see Fig.~\ref{fig-ele-bands}.
This is due to a sizeable splitting of the Te-derived CBM at $Q$  
which sets the CBM lower than the minimum at K, and thus the electronic band gap becomes indirect.

We now try to find out how general this phenomenon is by looking at the collection of band structures in 
Figures~\ref{fig-bands-cas-1} to \ref{fig-bands-cas-4}.
It is clearly seen that the intra-layer transitions in many MoSeTe/WSeTe bilayers become indirect. 
For CAS-1 and CAS-2, the lowest intra-layer transitions are from $K$ to $Q$, for the reason explained in the above paragraph. 
The situation is different in the CAS-3 bilayers: for most stackings, the lowest intra-layer transitions are indirect, but now from $\Gamma$ to $K$. 
This is because the electronic p orbitals of the Se atoms forming the junction have an overlap, which pushes the bands at the $\Gamma$ point in WSeTe and MoSeTe up and down, respectively (see Fig.~\ref{fig-bands-cas-3}(d) for the orbital contributions). The up-shift of the WSeTe valence maximum band at the $\Gamma$ point is so large that this point now becomes the band maximum, making the WSeTe band gap indirect (from $\Gamma$ to $K$). For the AA and AB-M stacking patterns of CAS-3, the direct band gap persists. 
As a feature common  to these metastable stackings, and in contrast to the more stable 2H stacking, the Se and Te atoms of MLs sit directly on top of each other. 
This enforces a larger $d_{\text{inter}}$ (see Tab.~\ref{table_thermal} ) and hence less overlap of the in-plane p-orbitals, resulting in a less dispersive valence band at $\Gamma$. 
These results are consistent with earlier DFT study on mixed bilayers (MoS$_2$/WSe$_2$), where direct and indirect band gaps were calculated for AA and AB stacking patterns, respectively~\cite{scirep-terrones-2013}. 
Also, given the direct band gap of around $1.0~\text{eV}$ for CAS-3:AA and CAS-3:AB-M bilayers, some applications can be envisaged in the infrared range for these types of bilayers. Finally, the CAS-4 bilayers differ from the other three as their band gap remains direct from $K$ to $K$, with the exception of the AA stacking pattern. 

\par
Table~\ref{table_electronics_splitting} shows 
the splitting of the valance band ($\Delta E^{\text{SOC}}_V$) at the high-symmetry point $K$ due to spin-orbit coupling. 
The $E^{\text{SOC}}_V$ values are larger than the conduction band splitting~\cite{PhysRevB-kosmider-14} since the d$_{x^2 -y^2}$ orbitals with the large magnetic quantum number of $m_l = \pm 2$ form the  valence band edge at K. 
The latter is similar to spin-orbit coupling in atomic physics: The orbital angular momentum couples with the spin to the total angular momentum of $m_l \pm \frac{1}{2}$. The atomic spin-orbit term in the  Hamiltonian is proportional to $Z^4/n^3$, where $Z$ and $n$ are nuclear charge and principal quantum numbers, respectively \cite{griffiths2018introduction}. As the W atom has larger $Z$ than Mo, $\Delta E^{\text{SOC}}_V$ is larger for W than for Mo. Also, the splitting is almost insensitive to the atomic sequence (CAS) and stacking. This suggests that such a band splitting originates to a large extent from the potential near the nuclei and the orbital angular momentum of the electrons, with just a very slight contribution from the electrostatic potential gradient across the bilayers. However, for the AB-stacked MoSeTe/WSeTe bilayer, notable (few tens of meV) deviations of the valence splitting are observed for both Mo and W atoms. For example,  $\Delta E^{\text{SOC}}_V$ for W is nearly $~50~\text{meV}$ larger when the CAS-2 bilayer has AB stacking.

\par
To conclude this Section, we would like to discuss the possibility of interlayer exciton formation in the Janus bilayers. 
While it is possible, though somewhat unlikely, to form an interlayer exciton directly in the process of light adsorption, the more likely path for creating an interlayer exciton is by optically exciting an electron plus a hole in one layer, followed by hopping of one carrier to the other layer. 
The advantage of creating an interlayer exciton lies in the fact that it is rather long-lived since the reduced overlap of the electron and hole wave function suppresses the matrix element for its radiative decay.  
In experimental work on the optoelectronic properties of hetero-bilayers with different junctions~\cite{apl-ufuk-22}, the possibility of interlayer exciton formation shows up as a quench of the photoluminescence signal. 
These observations assert the role of the conduction band offset of the constituting MLs in driving the excited electron into the adjacent layer. 
Since we already established type-II band alignment for the MoSeTe/WSeTe bilayer, electrons optically excited into the CBM of one ML may hop into the CBM of the adjacent monolayer, thereby releasing an energy $\Delta E_{\text{CM}}$ and getting captured. As a result of such electron transfer, interlayer excitons are formed in the bilayer. Since the CBM offset of the bilayer at the high-symmetry point $K$ is the driving force for the exciton formation, this offset largely determines the photoluminescence quench. 
Table~\ref{table_electronics} shows $\Delta E_{\text{CBM}}$ for the full collection of bilayers studied in the present work. Depending on the sequence of chalcogen atoms in the bilayer and the ML stacking, the magnitude of the CBM offset changes. Since the transition metal atoms make major contributions to the conduction band edges of the bilayer, the CBM offset 
is determined both by the energetic difference between the d shells of the transition metal atoms (with W 5d higher than Mo 4d) and a possible potential difference across the space between the two monolayers. 
The CBM offset becomes notable for CAS-4, in which {\em two different}  chalcogen species form the junction. This can be explained by the fact that, for CAS-4, the interface 
produces a potential drop over a short distance (almost $3.3$ to $4.0~\text{\AA}$), thus a notable dipole moment pointing in the $-z$ direction is generated, see Fig.~\ref{fig-dipole-aa-ab-stackings}. 
This potential drop points in the same direction as the intrinisic potential difference between the 4d and 5d orbitals of Mo and W, respectively, and thus both effects add up.
This CAS effect becomes slightly more pronounced when the chalcogen atoms at the interface sit on top of each other (AA and AB-M stacking, see Fig.~\ref{fig-sketch}), where the potential gradient in the gap between the two monolayers is oriented normal to the interface. For sufficiently large CBM offset, efficient charge separation can take place, yielding long-lived interlayer excitons. 
For this reason, a heterostructure bilayer can be a good candidate for photo-electrocatalysis applications which require long-lived excited electrons and holes to participate in the hydrogen ion reduction and oxidation of hydroxyl ions.

\begin{table*}[tbp]
\centering
\begin{tabular}{SSSSSSSS} \toprule
   {structure} & {parameter} & {AA}  & {AA-X} & {AA-M} & {AB} & {AB-X} & {AB-M} 
    \\ \midrule\midrule
{} & {$| k_{\text{R}}|~(\text{\AA}^{-1})$} & {0.123}  & {0.042} & {0.102} & {0.025} & {0.046} & {0.114}  \\
{CAS-1} & {$2E_{\text{R}}~(\text{meV})$} & {36.8}  & {3.0} & {17.6} & {3.2} & {8.7} & {49.2}\\
{SeMoTeSeWTe} & {$|\alpha|~(\text{meV \AA})$} & {298.5} & {143.1} & {345.7} & {125.1} & {192.6} & {431.7}  \\\midrule

{} & {$| k_{\text{R}}|~(\text{\AA}^{-1})$} & {0.106}  & {0.041} & {0.049} & {0.038} & {0.071} & {0.089}  \\
{CAS-2} & {$2E_{\text{R}}~(\text{meV})$} & {34.0} & {6.0} & {10.0} & {5.6} & {21.0} & {29.0}  \\
{SeMoTeTeWSe} & {$|\alpha|~(\text{meV \AA})$} & {318.7}  & {141.8} & {199.1} & {145.8} & {313.1} & {332.8}  \\\midrule

{} & {$| k_{\text{R}}|~(\text{\AA}^{-1})$} & {0.033}  & {0.004} & {0.008} & {0.046} & {0.008} & {0.047}  \\
{CAS-3} & {$2E_{\text{R}}~(\text{meV})$} & {6.0}  & {0.2} & {0.6} & {12.4} & {0.6} & {11.5}  \\
{TeMoSeSeWTe} & {$|\alpha|~(\text{meV \AA})$} & {163.9}  & {58.0} & {77.5} & {270.7} & {80.4} & {249.0} \\\midrule

{} & {$| k_{\text{R}}|~(\text{\AA}^{-1})$} & {0.134}  & {0.036} & {0.042} & {0.040} & {0.045} & {0.123}  \\
{CAS-4} & {$2E_{\text{R}}~(\text{meV})$} & {84.0}  & {8.0} & {9.0} & {9.0} & {9.6} & {40.0}  \\
{TeMoSeTeWSe} & {$|\alpha|~(\text{meV \AA})$} & {620.2}  & {212.0} & {230.2} & {223.8} & {212.6} & {337.7}  \\
 \bottomrule
\end{tabular}
\caption{Rashba energy $\Delta E$ (in eV), wave vector offset $| k_{\text{R}} |$ (in $\text{\AA}^{-1}$), and Rashba parameter $|\alpha |$ (in eV $\text{\AA}$) 
%due to Rashba band splitting 
at the high-symmetry point $\Gamma$.}
    \label{table_electronics_rashba}
    \end{table*}

\subsection{Rashba band splitting}

In the electronic band structures shown in Figures~\ref{fig-bands-cas-1} to \ref{fig-bands-cas-4}, which were calculated by including spin-orbit-coupling (SOC) in the Kohn-Sham Hamiltonian, a 
characteristic splitting of the topmost valence bands near the $\Gamma$ point is observed. 
In this Section, we will show that this splitting is a manifestation of the Rashba effect. 
In general, this effect may occur in a two-dimensional system in the presence of SOC if inversion symmetry is broken.\cite{Bychkov1984} 
Frequently studied systems are surface states \cite{prl-lashell-96,Bihlmayer2006} or semiconductor quantum wells. 
In both systems, the lack of inversion symmetry of the (envelope) wave function is decisive for the Rashba splitting. 
In semiconductor quantum wells, a proportionality of the Rashba parameter and the in-built electric field of the well has been demonstrated \cite{Basani_PhysRevB_55}, 
thus facilitating the modulation of the Rashba effect by an electrical gate voltage \cite{prl_nitta_97gaterashab,schapers1998}.  
While the effect of SOC is fully included in our first-principles treatment, it can be instructive to consider simple toy models where one starts from an electronic band structure disregarding SOC and then adds it in a second step by the Rashba-Bychkov Hamiltonian. 
While the structure of this Hamiltonian is inspired by the description of SOC in atomic physics, its usage in semiconductor physics is mostly empirically motivated. It reads 
\begin{equation}
H_{\mathrm{R}}=\alpha\left(\vec k \times \vec \sigma\right)_{z},
\label{eq:RashbaHam}
\end{equation}
where $\vec{\sigma} := (\sigma_x, \sigma_y, \sigma_z)$ is the Pauli vector, $\vec{k}$ is the electron wave vector and $\alpha$ 
is the so-called Rashba parameter. 
However, one should keep in mind that the above considerations only hold on a model level, and it needs to be checked if and to what extent it is useful to analyse results from first-principles calculations. 
When the above Rashba-Bychkov Hamiltonian is added to the Hamiltonian of a quasi-free particle with effective mass $m$, the electronic bands take the form 
\begin{equation}
\varepsilon_{\pm}(k)=\frac{\hbar^{2} k^{2}}{2 m} \pm \alpha k=\frac{\hbar^{2}}{2 m}\left(k \pm k_{\mathrm{R}}\right)^{2}-E_{\mathrm{R}},
\label{eq:eps_rashba}
\end{equation}
where $k_{\mathrm{R}}$ and $E_{\mathrm{R}}$ are the momentum offset and Rashba energy, respectively. 
In practice, we start from our DFT+SOC calculations and fit the top-most valence bands around the high-symmetry point $\Gamma$ to this form.  
From the fit parameters, we calculate the Rashba parameter 
$ \alpha = 2E_{\text{R}}/k_{\text{R}} $
as a simple measure of the strength of the Rashba effect. 
We note that these parameters are accessible to experimental studies via spin- and angle-resolved photoemission 
spectroscopy (Spin-ARPES)~\cite{prl-hochstrasser-02,prl-lashell-96,natcomm-feng-2019-rashba}. 
For application in a spin transistor, large $E_{\text{R}}$ and $k_{\text{R}}$ are desirable for 
achieving a significant phase offset for the spin channels. This makes effective spin-momentum locking feasible and paves the way for the fabrication of more efficient spintronics devices.

\begin{figure*}[ht]
\centering
\includegraphics[width=1.0\textwidth]{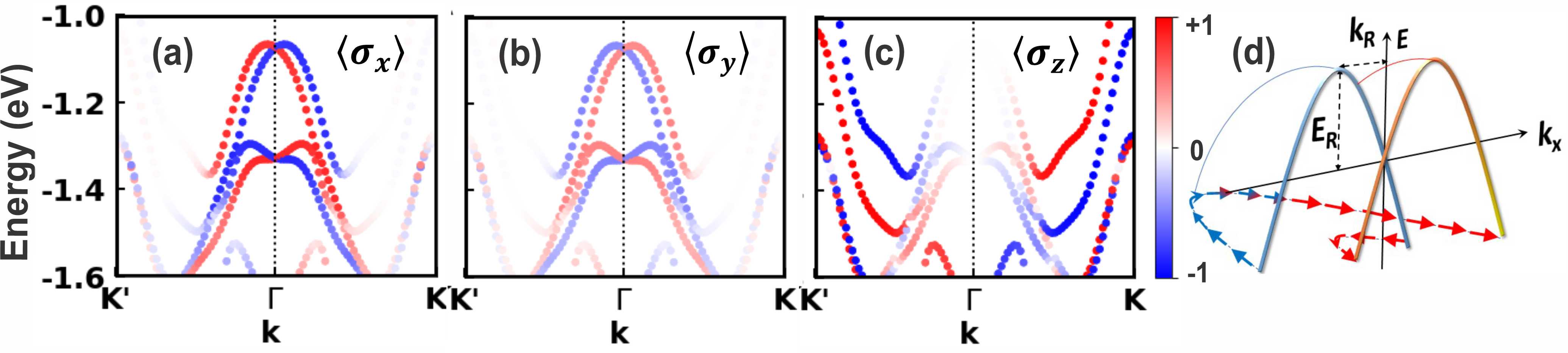}
\caption{Spin texture around the $\Gamma$ point for different chalcogen atom sequence CAS-2 and stacking pattern AB. The expectation values of the Pauli matrices along the $x$, (a) $y$ (b), and $z$ (c) axes. (d) The schematics of Rashba-type band splitting and shifting. }
\label{fig-ele-spin-texture-all-comps}
\end{figure*}

On the basis of the toy model eq.~(\ref{eq:RashbaHam}), one can show that the spin eigenstates have only in-plane components and that the orientation of the spin eigenvectors circles around the $\Gamma$-point of the Brillouin zone. 
For isotropic bands, these loops become circles, and the spin eigenvectors are tangential to the $\vec k$-vector at each point of the circumference. 
Due to the splitting given by eq.~(\ref{eq:eps_rashba}), there is an inner and outer loop, and the spins circulate clockwise or counterclockwise on each of the loops. 
In a one-dimensional cut through the Brillouin zone, e.g. along the $k_x$-axis, the split bands show up as two parabolas, see Fig.~\ref{fig-ele-spin-texture-all-comps}. 
Associated with each parabolic branch, there is one direction of the spin pointing either out of or into the plane of the plot. 
This can be understood by noting that the in-plane components of electron momentum $k_{x,y}$ enter linearly into the Hamiltonian $H_{\text{R}}$. Consequently,  the Bloch wave functions take the form of 
$$ |\Psi_{\pm}\rangle = \frac{1}{2\pi\sqrt{2} }
e^{i \vec{k}_{\|} \cdot \vec{r}_{\|}} \left(\begin{array}{l}i e^{-i \varphi / 2} \\ \pm e^{i \varphi / 2}\end{array}\right) \, ,
$$ 
where $\varphi$ is the angle of $\vec{k}_{\|}$ with an in-plane unit vector. 
As a result, we have the $\langle\Psi_{\pm}|\sigma_{x,y,z} |\Psi_{\pm}\rangle = (\pm\cos\varphi, \mp\sin \varphi, 0)$. 

To probe if this type of spin texture characteristic of the Rashba effect is also found in the results of our DFT calculations, we calculated the expectation values of the Pauli matrices $\sigma_{x,y,z}$ for the Bloch states obtained from the FHI-aims calculation. 

\begin{figure}[ht]
\centering
\includegraphics[width=0.50\textwidth]{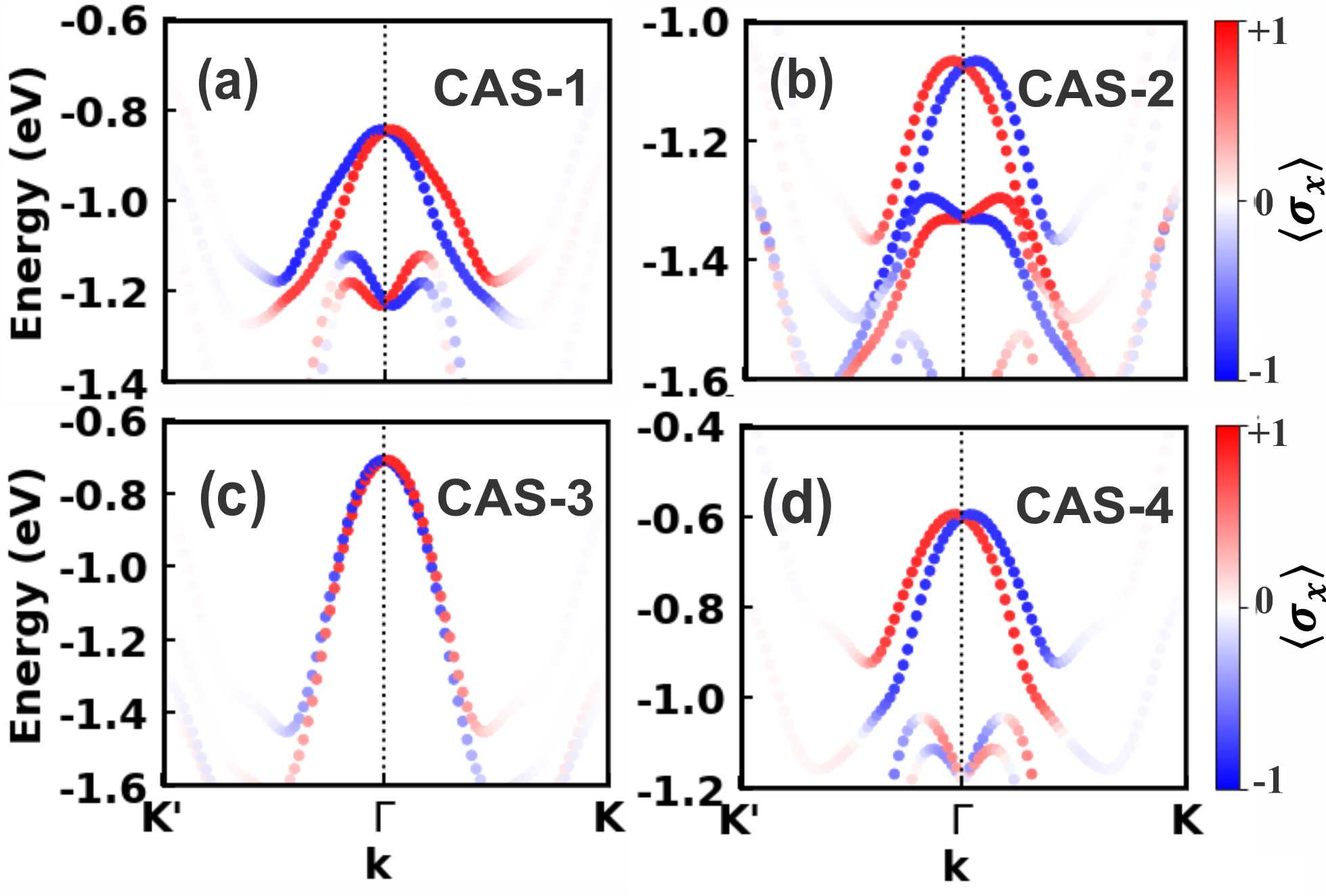}
\caption{Spin expectation value $\langle \sigma_x \rangle$ around the $\Gamma$ point for different chalcogen atom sequences CAS-1 (a), CAS-2 (b), CAS-3 (c), and CAS-4 (d) and stacking AB. The vertical color bar shows the expectation value of the Pauli matrix along the $x$ axis.}
\label{fig-ele-spin-texture}
\end{figure}

In Fig.~\ref{fig-ele-spin-texture-all-comps}, the expectation values 
of Pauli matrices are shown in color-coded form for top-most valence bands along the $k$-path $\text{K}'\rightarrow \Gamma \rightarrow \text{K}$ 
for bilayer CAS-2:AB. 
It is seen that $\left< \sigma_{x,y}\right>$ are non-zero and $\left< \sigma_{z} \right> \sim 0$ 
around the $\Gamma$-point for the two valence bands. 
For points of the $k$-path further away from $\Gamma$, the $\left< \sigma_{x,y}\right>$ get diminished. 
Also, one can decipher that spin vectors are oriented clockwise on the inner parabolic branch and counterclockwise 
on the outer branch, as expected from the solution of the Bychkov-Rashba Hamiltonian. 
This sense of rotation is found to depend on the chalcogen atom sequence. This can be observed from Fig.~\ref{fig-ele-spin-texture} 
that displays the color-coded expectation value of the $\sigma_x$ Pauli matrix around the $\Gamma$ point for the MoSeTe/WSeTe bilayers. 
While in CAS-1 the uppermost band colored in blue is shifted toward K' and the band colored red is shifted toward K (see Fig.~\ref{fig-ele-spin-texture}(a)), the opposite shifts are observed in CAS-4 (Fig.~\ref{fig-ele-spin-texture}(d)).  
In the CAS-2, CAS-3, and CAS-4 bilayers, the valence band is localized on the WSeTe ML, while in CAS-1 it belongs to the MoSeTe ML. In the CAS-2 and 
CAS-4 bilayers, the electric dipole moment of WSeTe ML is directed upward, so the spinors 
undergo rotations of the same sense around the $\Gamma$ point. In contrast, the spinor rotates in the opposite sense for 
CAS-1, where the electric dipole moment vector of the MoSeTe ML is pointing down. 

\par
Already when discussing the valence band splitting $\Delta E_{V}^{\text{SOC}}$ in Table \ref{table_electronics_splitting}, we pointed out that 
SOC, being a relativistic effect, strongly increases with the nuclear charge $Z$ of the atom. 
Analogously, we expect that the material-specific constant $\alpha \propto Z^4$, i.e. one (of several) factors governing the Rashba effect is larger for heavy atoms compared to light ones.   
Since both transition metal and Se atomic orbitals contribute to the valence band edge at the $\Gamma$ point, the $d_{z^2}$ and $p_z$ orbitals (or their relativistic counterparts) play important roles in the splitting. 
A previous theoretical study suggests that the contribution to the SOC is larger from $d_{z^2}$ than from $p_z$~\cite{prb-yao-17}. 
Analogously, a stronger Rashba-type splitting (greater Rashba parameter $\alpha$) is expected for the monolayer containing the heavier atoms. 
Putting aside a small difference in electric field strength between the MoSeTe and WSeTe MLs, larger Rashba-type SOC splitting is expected for WSeTe than for MoSeTe (since W is heavier than Mo), namely greater Rashba $k$-offset ($k_{\text{R}}$) and energy difference ($E_{\text{R}}$). 
Indeed, the Rashba $k$-offset and Rashba energy differences are $0.146~\text{\AA}^{-1}$ and $42.2~\text{meV}$ ($\alpha = 577.9~\text{meV}\text{\AA}$) for an isolated Janus WSeTe ML. 
For an isolated MoSeTe ML, $k_{\text{R}}$ and $E_{\text{R}}$ are $0.200~\text{\AA}^{-1}$ and $49.1~\text{meV}$ ($\alpha = 490.8~\text{meV}\text{\AA}$), respectively.

By building heterostructures from Janus monolayers, one may tune the Rashba effect and obtain a suitable Rashba parameter $\alpha$, which could be used for more effective locking of electron spin. 
The Rashba $k$-offset, energy difference, and $\alpha$ of the bilayers studied in the present work, obtained from fitting eq.~(\ref{eq:eps_rashba}) to the DFT+SOC band structures, are listed in Table~\ref{table_electronics_rashba}.

\par
It is seen that the values of $k_{\text{R}}$ and $E_{\text{R}}$ depend strongly on the chalcogen atomic sequence as well as on the stacking. For the stacking patterns AA and AB-M 
(when chalcogen atoms forming the junction sit on top of each other)  
the Rashba effect becomes strongest, i.e. larger $k$-offsets and energy differences occur. In Table~\ref{table_electronics_rashba}, the special role played by the AA and AB-M stacking patterns is clearly seen, and the variation with the chalcogen atom sequence is shown. 
It is noteworthy that the Rashba parameters become larger for CAS-1 and CAS-4 than for CAS-2 and CAS-3, see Table~\ref{table_electronics_rashba}. 
Also, the stark difference of the Rashba splitting between the CAS-2 and CAS-3 can be attributed to the chemical nature of their junctions. 
More explicitly, one can note that 
both in CAS-2 and CAS-3 the Se orbitals contribute most to the valence band top at the $\Gamma$-point. In CAS-2, the Se atoms sit at the upper and lower (outer) surface of the bilayer. However,  
in the CAS-3 bilayer, where Se atoms form the junction. 
Therefore, if one disregards the different chemical nature of W and Mo, there is still an approximate local symmetry near the junction. 
We suggest that this is the reason why the Rashba splitting, which relies on symmetry breaking, is very small in CAS-3.

For a few stacking patterns (CAS1:AB-M and CAS-4:AA), the Rashba parameters can be close to or even larger than the parameters of isolated monolayers. Therefore, a well-selected stacking pattern may provide a bilayer structure with suitable Rashba parameters for specific applications that require the generation of a well-tuned spin current. One may find merit in the variation of atomic order and stacking as suitable tools for fine-tuning the Rashba parameter. 
This is one central aspect pursued when selecting a suitable material for Rashba-based devices~\cite{jacs-gupta-21}, i.e. the ability to modulate the Rashba energy, $k$-offset, and $\alpha$ to get the most suitable Rashba effect.

\section{Conclusions} \label{Conclusions}
In summary, we have studied the thermodynamic stability, vibration spectrum, and electronic structure of MoSeTe/WSeTe bilayers with different chalcogen atom sequences and stacking patterns. To treat the van der Waals interactions between the layers, we have used both the pairwise additive approach of Tkatchenko-Scheffler and a many-body dispersion technique developed by Tkatchenko {\it et al}. Our results suggest that the many-body dispersion treatment 
based on the susceptibility of the interacting electronic system is capable of providing clear trends for the energy gain upon bilayer formation and the bilayer distance.

By analyzing the calculated results for the vibration spectrum we identify low-frequency Raman-active modes of rigid gliding or normal-to-layer motion, namely the inter-layer shear and layer-breathing modes, respectively. These can act as fingerprints of each atomic sequence and stacking orientation. Our theoretical results are in good agreement with low-frequency Raman spectroscopy measurements on hetero-bilayers and bring about further details on the origin of these types of vibration for each stacking.

Our results suggest that the atomic sequence plays a pivotal role in determining the electronic structure of the bilayer: {\it technologically} important electronic properties such as the type of the electronic band gap, conduction band offset,  as well as 
the band splitting at the $K$ ($K'$) points depend on the type of bilayer junction. Also, the strength of the Rashba effect is largely tied to the atomic sequence and the stacking. For some monolayer stackings, the Rashba parameters are close to or even greater than those of monolayers. Finally, it is noted that the ability to tune the Rashba band splitting as well as the electronic band gaps (type and magnitude) through varying the atomic sequence and stacking pattern of bilayers can make such 2D systems relevant 
for opto-electronics and energy conversion applications.

\section*{Author Contributions}
H.M. performed the DFT calculations.
P.K. conceived and supervised the project.
The manuscript was written through the contributions of both authors.
\par
The authors declare not to have any competing financial interests.

%%%%%%%%%%%%%%%%%%%%%%%%%%%%%

\section*{Acknowledgements}
This work is unded by the Deutsche Forschungsgemeinschaft (DFG, German Research Foundation) through project B02 of Collaborative Research Center SFB1242 'Nonequilibrium dynamics of condensed matter in the time domain' (Project-ID 278162697). 
The authors also  gratefully acknowledge computing time provided by the Paderborn Center for Parallel Computing (PC2). 
The authors would like to thank Gabrielle Koknat from Duke University for help with part of the coding.


%\bibliography{Bibliography}
\end{document}